\documentclass[a4paper,11pt]{article}

\let\latexput\put
\usepackage{pictexwd}

\pdfoutput=1 

\usepackage{jheppub} 
\usepackage{amsmath}                    
\usepackage[bbgreekl]{mathbbol}

\usepackage{amsfonts,amssymb,amsthm,mathtools} 
\usepackage[T1]{fontenc} 

\usepackage{psfrag}
\usepackage{indentfirst} 
\usepackage{bm}          
\usepackage[usenames,dvipsnames]{xcolor}
\usepackage{braket}
\usepackage{bbm}
\usepackage{dsfont}
\usepackage{amssymb,amsfonts,latexsym,cancel}
\usepackage{ulem}
\usepackage{amsthm}    
\usepackage{wrapfig}
\usepackage{subfigure} 
\usepackage{rawfonts}
\usepackage{pictexwd}
\usepackage{soul}
\usepackage{url}
\usepackage{enumitem}
\usepackage{MnSymbol}
\usepackage{hyperref}
\usepackage{tensor}
\usepackage[titletoc,title]{appendix}
\usepackage{mathtools}
\usepackage{xcolor} 
\usepackage{scalerel}
\usepackage{graphicx}
\usepackage{todonotes}
\usepackage{upgreek}
 
\providecommand{\abs}[1]{\lvert#1\rvert}
\DeclareSymbolFontAlphabet{\mathbbm}{bbold}
\DeclareSymbolFontAlphabet{\mathbb}{AMSb}
\DeclareSymbolFontAlphabet{\mathbbl}{bbold}

\AtBeginDocument{}%

\makeatletter
\newcommand\rrule[3][0pt]{%
	\ifdim#2>#3\math@hrule[#1]{#2}{#3}\else\math@vrule[#1]{#2}{#3}\fi}
\newcommand\math@hrule[3][0pt]{%
	\gdef\mystery@factor{0.07}%
	\@tempdima=#3%
	\rule[#1]{0pt}{#3}
	\raisebox{.5\@tempdima+#1}{%
		\makebox[#2][l]{\kern-.5\@tempdima\@@mathrule{#2}{#3}}}%
}
\newcommand\math@vrule[3][0pt]{%
	\gdef\mystery@factor{0.0}%
	\@tempdima=#2%
	\rule[#1]{0pt}{#3}
	\raisebox{-.0\@tempdima+#1}{%
		\kern0.5\@tempdima%
		\rotatebox{90}{\kern-0.5\@tempdima\makebox[#3][l]{\@@mathrule{#3}{#2}}}%
		\kern0.5\@tempdima}%
}
\def\@@mathrule#1#2{%
	\@tempdimb=#2%
	\@tempdima=\dimexpr#1-\mystery@factor\@tempdimb
	\pdfliteral{%
		q []0 d %
		1 J 
		\strip@pt\@tempdimb\space w \strip@pt\@tempdimb\space 0 m %
		\strip@pt\@tempdima\space 0 l S Q }}
\makeatother

\makeatletter
\DeclareFontFamily{OMX}{MnSymbolE}{}
\DeclareSymbolFont{MnLargeSymbols}{OMX}{MnSymbolE}{m}{n}
\SetSymbolFont{MnLargeSymbols}{bold}{OMX}{MnSymbolE}{b}{n}
\DeclareFontShape{OMX}{MnSymbolE}{m}{n}{
<-6>  MnSymbolE5
<6-7>  MnSymbolE6
<7-8>  MnSymbolE7
<8-9>  MnSymbolE8
<9-10> MnSymbolE9
<10-12> MnSymbolE10
<12->   MnSymbolE12
}{}
\DeclareFontShape{OMX}{MnSymbolE}{b}{n}{
<-6>  MnSymbolE-Bold5
<6-7>  MnSymbolE-Bold6
<7-8>  MnSymbolE-Bold7
<8-9>  MnSymbolE-Bold8
<9-10> MnSymbolE-Bold9
<10-12> MnSymbolE-Bold10
<12->   MnSymbolE-Bold12
}{}

\let\llangle\@undefined
\let\rrangle\@undefined
\DeclareMathDelimiter{\llangle}{\mathopen}%
{MnLargeSymbols}{'164}{MnLargeSymbols}{'164}
\DeclareMathDelimiter{\rrangle}{\mathclose}%
{MnLargeSymbols}{'171}{MnLargeSymbols}{'171}
\makeatother

\def\wwedgee{{\setbox0\hbox{\ensuremath{\mathrel{\wedge}}}\rlap{\hbox to \wd0{\hss\hspace*{.6ex}\ensuremath\wedge\hss}}\box0}}

\newcommand{\www}{\mathrel{\wwedgee}}
\newcommand{\dd}{\mathchoice
	{\mathbbm{d}\rrule{.087ex}{1.605ex}\hspace*{0.15ex}} 
	{\mathbbm{d}\rrule{.087ex}{1.605ex}\hspace*{0.15ex}} 
	{\mathbbm{d}\rrule{.08ex}{1.125ex}\hspace*{0.15ex}}  
	{\mathbbm{d}\rrule{.06ex}{.8ex}\hspace*{0.15ex}}     
}

\renewcommand{\d}{\mathrm{d}}
\def\R{\mathbb{R}}

\def\ll{\left\langle}
\def\rl{\right\rangle}

\title{\boldmath Edge observables of the Maxwell-Chern-Simons theory}

\author[a, d]{J. Fernando Barbero G.,}
\author[b, d]{Bogar D\'{\i}az,}
\author[c,d]{Juan Margalef-Bentabol,}
\author[b,d]{and Eduardo J.S. Villase\~nor}

\affiliation[a]{Instituto de Estructura de la Materia, CSIC. Serrano 123, 28006 Madrid, Spain}
\affiliation[b]{Departamento de Matem\'aticas, Universidad Carlos III de Madrid, Avenida  de la Universidad 30, 28911 Legan\'es, Spain}
\affiliation[c]{Department of Mathematics and Statistics, Memorial University, St. John's, Newfoundland and Labrador A1C 5S7, Canada}
\affiliation[d]{Grupo de Teor\'{\i}as de Campos y F\'{\i}sica Estad\'{\i}stica. Instituto Gregorio Mill\'an (UC3M). Unidad Asociada al Instituto de Estructura de la Materia, CSIC, Serrano 123, 28006 Madrid, Spain}

\emailAdd{fbarbero@iem.cfmac.csic.es}
\emailAdd{bodiazj@math.uc3m.es}
\emailAdd{juanmargalef@mun.ca}
\emailAdd{ejsanche@math.uc3m.es}

\abstract{We analyze the Lagrangian and Hamiltonian formulations of the Maxwell-Chern-Simons theory defined on a manifold with boundary for two different sets of boundary equations derived from a variational principle. We pay special attention to the identification of the infinite chains of boundary constraints and their resolution. We identify edge observables and their algebra [which corresponds to the well-known $U(1)$ Kac-Moody algebra]. Without performing any gauge fixing, and using the Hodge-Morrey theorem, we solve the Hamilton equations whenever possible. In order to give explicit solutions, we consider the particular case in which the fields are defined on a $2$-disk. Finally, we study the Fock quantization of the system and discuss the quantum edge observables and states.}

\arxivnumber{}

\let\put\latexput

\begin{document}
\maketitle
\flushbottom

\section{Introduction}\label{sec_introduction}

The Chern-Simons (CS) theory in manifolds with boundaries is a very interesting model as pointed out by Witten in \cite{Witten} (see also \cite{elitzur1989remarks}). It plays a relevant role in condensed matter physics, in particular in the study of the integral and fractional Hall effects \cite{Wen1992,wen1995topological,Zee:1995avy,Witten:2015aoa}. The system obtained by adding the Maxwell and CS Lagrangians (MCS) describes important physical phenomena, among them the gap from the fundamental state and bulk elementary excitations \cite{Wen1992} and topologically massive spinor electrodynamics \cite{deser1982three,DESER1982}. An interesting feature of the CS and MCS dynamics is the appearance of edge excitations \cite{Balachandran:1994upto, Asorey2016, Asorey2013, Agarwal2017} and edge observables \cite{Bala1992,Banados1995,Bala2003,Donnelly:2016auv,Geiller:2017xad}. Edge excitations play a significant role to explain the transport properties of integer quantum Hall states \cite{wen2004quantum}, whereas in the case of the fractional Hall effect it is necessary to rely on a low energy effective theory, obtained by using the so-called hydrodynamical approximation, in which the Kac-Moody algebra plays a central role. This effective theory turns out to be given by an Abelian CS Lagrangian \cite{wen2004quantum,fradkin2013field}. In this context the edge states correspond to classical solutions to the effective field equations which are essentially supported on the boundary. Edge observables also appear in  general relativity \cite{Balachandran:1994up}. For instance, the Einstein-Maxwell-Chern-Simons theory has played a relevant role in the study of $(2+1)$-dimensional black holes  \cite{Andrade2005}. In this case, the  black hole horizon acts as a spacetime boundary. 

Hamiltonian methods are important, among other things, as the starting point for canonical quantization. In the context of the MCS model in manifolds with boundaries, these have been discussed by a number of authors \cite{Bala1994,Park1999,Blasi2010}. In the particular case of a disk, the identification of edge observables and their algebra (whose relevance on general grounds was already pointed out in \cite{Witten}) has been highlighted in \cite{Bala1994}, as well as their role in the Dirac quantization of the system.

The Dirac analysis of field theories defined on manifolds with boundaries exhibits a number of interesting features, in particular with regard to the boundary dynamics (as defined by the action) and the role of boundary conditions. For instance, a characteristic phenomenon, which is often neglected, is the appearance of infinite chains of boundary constraints, which are necessary for the dynamical consistency of the model. In the case of the scalar field, it is well known that these chains of constraints play an important role related to the smoothness of the solutions to the field equations \cite{brezis2010}. From a practical point of view, the best way to implement the Dirac algorithm for field theories with boundaries is the geometric approach discussed in \cite{Diracnos} [or a similar one based on the Gotay-Nester-Hinds (GNH) method \cite{GNH1,Barbero_G_2014,margalef2018}]. 

In the present work we give a general discussion of the Hamiltonian formulation of the MCS model on a compact manifold $\Sigma$ with boundary. We consider two different situations which are taken into account by adding a surface term to the Lagrangian proportional to a non-negative coupling constant $\lambda^2$, which may be equal or different from zero. As pointed out in \cite{Bala1994}, this parameter has a physical interpretation in the case where the manifold $\Sigma$ (actually a disk) is surrounded by a superconductor. By relying on a geometric version of Dirac's method \cite{Diracnos}, we find \textit{all of the constraints}, including the often neglected chains of constraints at the boundary. We then discuss the edge observables, their evolution, and their algebra. By using the Hodge-Morrey theorem, we solve the Hamilton equations of motion, characterize in a precise way the reduced phase space, and give a concrete description both of the Hamiltonian and the edge observables. These results lead to a straightforward quantization of the model in the reduced phase space. In order to make contact with the results of \cite{Bala1994}, we consider in detail the case in which $\Sigma$ is a disk, in particular we give the full solution to the Hamilton equations for the $\lambda=0$ case, the edge observables, and the solutions that play the role of edge states. In the $\lambda\neq0$ case, the concrete description of the reduce phase space is not direct. At any rate, we have been able to complete the resolution of the field equations for the pure Maxwell case. It is important to note that we have not used any gauge fixing but, rather, given explicit descriptions of the relevant reduced phase spaces. 

The structure of the paper is the following. After this introduction, in Section \ref{sec:2}, we use the Abelian CS model to illustrate some issues relevant for the study of the MCS theory. In Section \ref{sec:3} we present the Lagrangian and Hamiltonian analysis of the MCS model for two natural sets of boundary conditions. Whenever possible, we solve the resulting Hamilton equations of motion together with all the constraints, and use these solutions to carry out the Fock quantization of the theory. Furthermore, we discuss the classical and quantum edge observables of the model. Finally, we end the paper with our conclusions in Section \ref{sec:conclu} and two appendixes. In the first one we solve an eigenvalue problem for the $\delta \mathrm{d}$ operator and in the second we give the relevant details about the derivation of the infinite chains of boundary constraints.

\section{Abelian Chern-Simons}\label{sec:2}

In this section, we use a simple example to illustrate some features of gauge theories defined on manifolds with boundary that we exploit in the next section for the models that we study in the paper. Let $\Sigma$ be a two-dimensional compact manifold with boundary and $M=\mathbb{R}\times \Sigma$. The action 
\begin{equation}
S_{\mathrm{CS}}( \bm{A} ) =  \int_{M} \bm{A} \wedge \mathrm{d} \bm{A} \,,  \label{eq:CSNN}
\end{equation}
where $\bm{F} = \mathrm{d} \bm{A}$ is the curvature of a connection 1-form $\bm{A}$, defines the Abelian Chern-Simons model. Notice that the field space is $\mathcal{F}=\Omega^1\left(M\right)$, and that we have not introduced any other condition in its definition (this is of the utmost importance when deriving the Euler-Lagrange equations). The field equations are
\begin{subequations}
\begin{align}
    \mathrm{d} \bm{A} &=0\,, \label{becs}\\
 \jmath^*_{\partial} (  \bm{A}) &=0\,, \label{nbccs}
\end{align}
\end{subequations}
where $\jmath_\partial:\partial M \hookrightarrow  M$ is the natural inclusion of the the boundary $\partial M:=\mathbb{R}\times \partial \Sigma$ in $M$, and $\jmath_\partial^*$ denotes the corresponding pullback. The bulk equation \eqref{becs}  tells us that the connection $\bm{A}$ must be flat, and \eqref{nbccs} are boundary conditions of the Dirichlet type. Notice that other boundary conditions---that can be included in the definition of the field space---may be compatible with the action principle \eqref{eq:CSNN}. This will be made clear as soon as we perform the $2+1$ decomposition, as we discuss now.

To this end, we consider the $2$-surfaces, $\Sigma_t:=\{t\}\times\Sigma$, of constant $t$, diffeomorphic  to $\Sigma$, where $t$ is the scalar function defined on $M$  as $t:\mathbb{R}\times\Sigma\rightarrow\mathbb{R}:(\tau,p)\mapsto \tau$. For $p\in\Sigma$, the vectors tangent to the curves $c:\mathbb{R}\rightarrow M: t\mapsto  (t,p)$ define a vector field ${\bf{t}}$ satisfying the condition  $\pounds_{{\bf{t}}} t=1$. In the following, we  use nonbold fonts for the objects living on $\Sigma$ to distinguish them from those defined on $M$. Using  the standard decomposition of the connection, the configuration space is $\mathcal{Q}=\{ (A_t, A)\,|\, A_t \in C^{\infty} (\Sigma) , A \in \Omega^1(\Sigma)\}$, and the action \eqref{eq:CSNN} can be written as
\begin{equation}
S_{\mathrm{CS}} (A_t, A) = \int_{\mathbb{R}} \d t  \int_{ \Sigma}  \left(\pounds_{{\bf{t}}} A \wedge A+ A \wedge  \d A_{t} +  A_{t} \d A \right)\,. \label{DCS}
\end{equation}
By demanding the stationarity of \eqref{DCS}, we obtain
\begin{equation}
\int_{\mathbb{R}} \d t  \int_{ \Sigma} 2 \left(a_{t} \d A - a \wedge \left( \pounds_{{\bf{t}}} A -\d A_{t} \right) \right) + \int_{\mathbb{R}} \d t  \int_{\partial \Sigma} \imath_\partial^*\left( A_{t} a - a_{t} A \right)=0\,, \quad \forall (a_t,a)\,, \label{vcs}
\end{equation}
where $(a_t,a)$ denotes the variations of $(A_t,A)$, $\imath_\partial:\partial \Sigma \hookrightarrow  \Sigma$ is the natural inclusion, and $\imath_\partial^*$ its pullback. We get $F:=\d A=0$ and $\pounds_{{\bf{t}}} A -\d A_t=0$ in the bulk, as expected. The vanishing of the boundary term implies $\imath_\partial^*\left(A_{t} \right)=0= \imath_\partial^*\left(A \right)=0$. It is important to notice that, in principle, we can include the following conditions (which are independent of the shape of the boundary) in the definition of the configuration space:
\begin{enumerate}[label=(\roman*)]
    \item $\imath_\partial^*\left(A \right)=0$, which leads to the condition  $\imath^*_\partial(a)=0$  on the variations of $A$. Then \eqref{vcs} implies that $A_t$ is arbitrary at the boundary,  or\label{qpc2}
    \item $\imath_\partial^*\left( A_t \right) =0$, which leads to $\imath_\partial^*\left( a_t \right) =0$. Then \eqref{vcs} implies that $A$ is arbitrary  at the boundary\,. \label{qpc}
\end{enumerate}
Both (i) and (ii) are compatible with the action \eqref{DCS} in the sense that the boundary term in \eqref{vcs} vanishes as a consequence of them. The conditions \ref{qpc2} trivialize the edge observables that we will construct below; therefore we will work with \ref{qpc} from now on. We must mention that if we are interested on a particular boundary (as in the next section), we can take advantage of its particular shape to write specific boundary conditions. For instance, if we consider the particular case in which the boundary is a disk of radius $R$, then $a= a_r \d r +a_\theta \d \theta$ and the boundary term $\imath_\partial^*\left(A_{t} a - a_t A \right)$ can be written as  $\left(A_{t} a_\theta - a_{t} A_\theta \right)|_{r=R}$. This term vanishes if we introduce in the definition of the configuration space the conditions $A_{t}|_{r=R}=A_\theta|_{r=R}$ or $A_{t}|_{r=R}=-A_\theta|_{r=R}$.

We must remark that adding boundary terms to the action may change the boundary conditions as a consequence of the boundary dynamics determined by the action. For example, the conditions \ref{qpc} 
can become part of the field equations (i.e., we do not have to put them \textit{a priori} in the definition of the configuration space) if we add the following boundary term
\begin{equation}
   \int_{\mathbb{R}} \d t  \int_{\partial \Sigma} \imath_\partial^*\left( A_{t} A\right) \label{abtcs}
\end{equation}
to the Chern-Simons action \eqref{DCS}.  For this reason,  in the following we will work with the action
\begin{align}\begin{split}
S_{\mathrm{CSb}}(A_t, A) &=     \int_{\mathbb{R}} \d t \int_{ \Sigma}  \left( \pounds_{{\bf{t}}} A \wedge A+ A \wedge  \d A_{t} +  A_{t} \d A \right)+ \int_{\mathbb{R}} \d t  \int_{\partial \Sigma} \imath_\partial^*\left( A_{t} A\right)\\
&= \int_{\mathbb{R}} \d t \int_{ \Sigma} \left(  \pounds_{{\bf{t}}} A \wedge A+  2 A_{t} \d A \right)\,,
\end{split} \label{CS2}
\end{align}
whose boundary equations are just  \ref{qpc}. We must mention that \eqref{abtcs} [and, hence, \eqref{CS2}] is adapted to the foliation $\mathbb{R}\times \Sigma$, and it cannot be written in a covariant ``spacetime'' form.

We summarize now the main results of the Hamiltonian analysis of the action \eqref{CS2} using the geometric implementation of Dirac’s algorithm  discussed in \cite{Diracnos, HKnos} (similar information can be obtained by using the GNH method \cite{GNH1, Barbero_G_2014, barbero2021b,Valle-Marc2021, margalef2018}). The submanifold in phase space where the dynamics takes place is 
\begin{equation*}
C:=\left\{  \left( A_{t}, A, {\bf p}_{t}, {\bf p} \right) \in T^*\mathcal{Q} :  {\bf p}_{t}(\cdot)=0,\, {\bf p}(\cdot)- \int_{\Sigma} (\cdot) \wedge A = 0, \,    \d A=0, \, \imath_\partial^*(A_{t})=0   \right\}\,,
\end{equation*}
and the components of the Hamiltonian vector field are
\begin{eqnarray}\label{HVFCSG}
 X_{A\, t}=  \mu_t\,, &\qquad&\displaystyle {\bf X}_{{\bf p}\, {t}}   \left(\cdot\right)= 0\,, \nonumber\\
 X_{A}  =  \d A_{t}\,, &\qquad& {\bf X}_{{\bf p}}  \left(\cdot\right)=    \int_{\Sigma} \left(\cdot\right) \wedge \d A_{t}\,,
\end{eqnarray}
with the Dirac multiplier $\mu_t\in C^{\infty} (\Sigma)$  vanishing at the boundary, i.e., $ \imath^*_{\partial} \left( \mu_t \right)=0$, but otherwise arbitrary. 

The presence of the arbitrary function of time $\mu_t$ in the Hamiltonian vector field \eqref{HVFCSG}, which implies that $A_t$ is arbitrary, can be immediately interpreted as the Abelian gauge symmetry $A \mapsto A+\d \epsilon$ with $\epsilon \in C^{\infty} (\Sigma)$ and $\imath_\partial^*(\epsilon)=0$.
\subsection{Classical edge observables}\label{ceoCS}

Let us construct the so-called classical edge observables \cite{Bala1992}. Given any $\Lambda \in C^{\infty} (\Sigma)$, we define 
\begin{equation}\label{eq:obsCS}
Q_{\Lambda}(A):= \int_{\Sigma} \d \Lambda \wedge A \,.
\end{equation}
First, notice that the functions \eqref{eq:obsCS} are invariant under the gauge transformations of the theory, $A'= A+\d \epsilon$,  because
\begin{equation*}
Q_{\Lambda}(A')= \int_{\Sigma} \d \Lambda \wedge A' =  Q_{\Lambda}(A)- \int_{\partial\Sigma} \imath_\partial^* \left( \epsilon  \d \Lambda\right) = Q_{\Lambda}(A) \,,
\end{equation*}
where we have used $\imath_\partial^*(\epsilon)=0$ [which is a consequence of $\imath_\partial^*\left( A_t \right) =0$]. Therefore, the functions \eqref{eq:obsCS} are {\it{observables}}. Second, on the constraint submanifold defined by the condition $\d A=0$, they satisfy
\begin{equation*}
Q_{\Lambda}(A)= \int_{\Sigma} \d \Lambda \wedge  A =  - \int_{\Sigma}  \Lambda \d A+\int_{\partial \Sigma} \imath_\partial^* \left( \Lambda A \right)= \int_{\partial \Sigma}  \imath_\partial^*\left( \Lambda A \right) \,.
\end{equation*}
This means that, for a given solution $A$, the functions \eqref{eq:obsCS} are characterized by the value of $\Lambda$ on the boundary, $\imath_\partial^*(\Lambda)$; this is the reason they are called edge observables.

Using the Hamiltonian vector field \eqref{HVFCSG}, we can calculate the evolution of the $Q_{\Lambda}(A)$
\begin{equation}
\dot{Q}_{\Lambda}(A)= \int_{\Sigma} \d \Lambda \wedge X_A 
= \int_{\partial \Sigma}  \imath_\partial^*\left( \Lambda  \d A_{t} \right) =0 \,,
\end{equation}
where we have made use of $\imath_\partial^*\left(A_{t}\right)=0$. As we can see, the $Q_{\Lambda}(A)$ are constants of motion. Another interesting aspect of these edge observables is related to their algebraic properties. If one uses the Poisson brackets of the full phase space, the edge observables satisfy  $\{Q_{\Lambda_1}(A), Q_{\Lambda_2}(A) \}=0$. However, if one is interested in quantization, the presence of second-class constraints prevents us from doing this. Instead, the relevant Poisson algebra of the edge observables must be computed with the Poisson brackets $\{\cdot,\cdot\}_{\mathrm{PB}}$ defined by the pullback of the canonical symplectic form onto the phase space submanifold defined by
\[
 {\bf p}(\cdot)-  \int_{\Sigma} (\cdot) \wedge A = 0\,. 
\]
By doing this we find
\begin{equation}
\label{KacMoodyCS}
\{Q_{\Lambda_1}(A), Q_{\Lambda_2}(A)\}_{\mathrm{PB}} = \frac{1}{2} \int_\Sigma  \d \Lambda_1 \wedge \d \Lambda_2= \frac{1}{4} \int_{\partial \Sigma}  \imath_\partial^* \left( \Lambda_1 \d \Lambda_2- \Lambda_2 \d \Lambda_1 \right)\,. 
\end{equation}
The same result can be obtained by using Dirac brackets \cite{Ferrari1997}. When  $\partial\Sigma \cong \mathbb{S}^1$, these observables generate a $U(1)$ Kac-Moody algebra \cite{Segal} localized on $\partial \Sigma$. The relevance of the loop group $LU(1)$ (and in general of $LG$) to the treatment of the CS theory was pointed out by Witten in his celebrated 1989s paper \cite{Witten}. In particular, for the edge observables of the pure CS theory on the disk, the algebra \eqref{KacMoodyCS} was found in \cite{Bala1992}.

Finally, notice that the construction of these edge observables is based on the first-class constraints of the theory, in this case $\d A=0$, which guarantee their gauge invariance and show that they vanish in the bulk. We must also remark the importance of the condition $\imath_\partial^*\left(A_{t}\right)=0$. The procedure discussed above suggests a way to construct edge observables for other gauge theories. However, we must say that they may or may not exist as well-defined operators in a fully quantized theory.

\section{The Maxwell-Chern-Simons Model}\label{sec:3}

The main purpose of the present paper is to study the Maxwell-Chern-Simons model. For a three-dimensional manifold $M=\mathbb{R}\times\Sigma$ this is defined by the action
\begin{equation}
S_{\mathrm{MCS}}(\bm{A}) =  \int_{M} \Big(\alpha \bm{F} \wedge \star \bm{F}+\beta \bm{A} \wedge \bm{F}\Big),\label{MCS3}
\end{equation}
where $\alpha$ and $\beta$ are  nonzero real constants, $\bm{F} = \mathrm{d} \bm{A}$ is the curvature of the three-dimensional connection 1-form $\bm{A}$, and $\star$ is the Hodge dual in $M$ with respect to the Minkowski metric with signature $(-,+,+)$. The field space is $\mathcal{F}=\Omega^1\left(M\right)$. In the following, we will work with the Lagrangian
\[\begin{array}{cccc}
L: &  T\mathcal{Q}:= T\left( C_0^{\infty} (\Sigma)\times \Omega^1(\Sigma) \right)  & \longrightarrow & \R \\
 &          \textrm{v}=((A_{t},A ),(v_{t},v))  & \longmapsto & L(\textrm{v} )
\end{array} \]
given by
\begin{align}
L(\textrm{v})=&  \int_{\Sigma} \big( -\alpha \left( v- \mathrm{d} A_{t} \right) \wedge \ast  \left( v- \mathrm{d} A_{t} \right)  + \alpha \left( \ast \mathrm{d} A\right) \mathrm{d} A  +\beta  \left( v- \mathrm{d} A_{t} \right)  \wedge A+\beta A_{t} \mathrm{d} A \big) \nonumber\\
&+\int_{\partial \Sigma}  \imath_\partial^*\left( \lambda^2  A \, \iota_\nu \ast A   \right)
\,, \label{eq:LagraMCSB}
\end{align}
where we have considered a foliation by inertial observers and, as in the previous section, $\imath_\partial$ is the natural inclusion of $\partial \Sigma$ in $\Sigma$, $\imath_\partial^*$ its pullback, and $\ast$ is the Hodge dual in $\Sigma$ with respect to the induced metric. The  bulk terms in \eqref{eq:LagraMCSB} correspond to the $2+1$ decomposition of the action \eqref{MCS3} (which is performed by introducing the same geometrical objects as in the Chern-Simons case) and we have added a boundary term. In that term $\lambda$ is a function on $\partial\Sigma$, $\nu$ is the outer unit normal to the boundary, and $\iota_\nu  \vartheta$ denotes the interior product (contraction) of $\nu$ with the differential form $\vartheta$. Finally, notice that the notation $C_0^{\infty} (\Sigma)$ means that $\imath_\partial^*\left( A_t \right)=0$, i.e., we incorporate this boundary condition in the definition of the configuration space. 

The role of the boundary term in \eqref{eq:LagraMCSB} is to give 
\begin{equation*}
\imath_\partial^* \left( \{\alpha \ast \mathrm{d}  +  \lambda^2 \iota_\nu \ast\} A \right)=0\,,
\end{equation*} 
as boundary equations  naturally derived from the variational principle. Notice that the boundary term [and hence the action \eqref{eq:LagraMCSB}] is adapted to the foliation $\mathbb{R}\times \Sigma$ so it cannot be written in a covariant form. 
As we show below, some details of the Hamiltonian analysis strongly depend on $\lambda$. In fact, it is useful to treat the cases $\lambda=0$ and $\lambda\neq0$ separately. Finally, we must mention that the added boundary term is compatible with the gauge symmetries of the theory (which we will get below), in particular $A \mapsto A+ \d \epsilon$ with $\imath_\partial^* \left(\epsilon\right)=0$.

\subsection{Hamiltonian formulation}\label{Hf}

In this section, we give the relevant steps to obtain the Hamiltonian formulation of the model defined by the Lagrangian \eqref{eq:LagraMCSB} using the geometric version of the Dirac algorithm \cite{Diracnos}. If we take $\mathrm{v}, \mathrm{w}$ in the same fiber of $T\mathcal{Q}$, $\mathrm{v} :=((A_t,A,),(v_{t},v))\,, \mathrm{w} :=((A_t,A),(w_{t},w))\,,$
we get the fiber derivative, $F\!L:T \mathcal{Q}\to T^*\mathcal{Q}$,
\begin{equation}
\displaystyle \ll \boldsymbol{p} | \textrm{w} \rl =\left\langle F\!L\left(\textrm{v}\right)|  \textrm{w} \right\rangle=\int_{\Sigma} w \wedge  \ast \left(-2\alpha\left( v- \mathrm{d}  A_{t} \right) -\beta  \ast A \right)\,. \label{df3}
\end{equation}
Boldfaced letters will be used to denote elements of the dual space, i.e.,  $\boldsymbol{p}\in \left(C_0^{\infty} (\Sigma)\times \Omega^1(\Sigma) \right)^*$. Writing $\boldsymbol{p}:=\left( {\bf p}_{t}, {\bf p}\right)$, we read the momenta from \eqref{df3}
\begin{equation}
{\bf p}_{t}(\cdot) \!=\boldsymbol{0}\,, \qquad {\bf p}(\cdot) \!= \displaystyle \int_{\Sigma} \cdot \wedge  \ast \left(-2 \alpha\left( v- \mathrm{d}  A_{t} \right) -  \beta \ast A \right)\,. \label{fm3}
\end{equation}
The energy function is given by
\begin{align}
E:= \left\langle F\!L\left(\textrm{v}\right)|  \textrm{v} \right\rangle-L =&  \int_{\Sigma} \big(  -\alpha\left( v+ \mathrm{d}  A_{t} \right)  \wedge  \ast \left( v- \mathrm{d}  A_{t} \right)- \alpha  \left( \ast \mathrm{d} A  \right) \mathrm{d} A +\beta \left(    \mathrm{d} A_{t} \wedge A -  A_{t}  \mathrm{d} A \right) \big) \nonumber\\
&- \int_{\partial \Sigma}  \imath_\partial^*\left(\lambda^2  A \, \iota_\nu \ast A   \right) \,. \label{eq:energy}
\end{align}
An easy way to write down the Hamiltonian is to represent the canonical momenta in terms of differential forms. Explicitly, taking advantage of the fact that the Hodge operator defines a scalar product, the momenta can be written as
\begin{equation*}
{\bf p}_{t}(\cdot) = \int_{\Sigma} \cdot \,  \ast p_{t}\,, \qquad  {\bf p}(\cdot) = \int_{\Sigma} \cdot \, \wedge  \ast p \,,
\end{equation*}
where  $p_{t} \in C^{\infty} (\Sigma)$ and $p\in \Omega^1( \Sigma)$. From \eqref{fm3}, we obtain
\begin{equation}
p_{t}=0 \,,  \qquad  p= -2 \alpha\left( v- \mathrm{d}  A_{t} \right) - \beta\, \ast A\,.  \label{eq:cm}
\end{equation}
Notice that the first equation in \eqref{eq:cm} is a primary constraint. Plugging \eqref{eq:cm} into \eqref{eq:energy} gives the Hamiltonian 
\begin{align}\label{Hamiltonian_MCSB}
H =& \int_{\Sigma}\left( \left(  \mathrm{d} A_{t} -\frac{1}{4 \alpha} \left( p+\beta \ast A\right) \right)\wedge \ast  \left( p+\beta \ast A\right)  - \alpha \left( \ast \mathrm{d} A \right)\mathrm{d} A  
+\beta \left(  \mathrm{d} A_{t} \wedge A -  A_{t}  \mathrm{d} A \right)\right) \nonumber\\
&-  \int_{\partial \Sigma}  \imath_\partial^*\left(\lambda^2  A \, \iota_\nu \ast A   \right) \,.
\end{align}
\subsubsection*{Dirac analysis in the bulk}
The constraints in the bulk are 
\begin{equation*}
p_{t}=0\,, \qquad  \delta \left( p-\beta \ast A \right)=0\,,    
\end{equation*}
where $\delta$ is the codifferential defined as $\delta=-\ast \mathrm{d} \ast$ when acting on forms of any order. In the previous computations, we have used $\ast \ast =(-1)^{k(2-k)}=(-1)^{k}$ on $k$-forms.

The components of the Hamiltonian vector field are
\begin{equation}\label{VF33} \begin{alignedat}{2}
 X_{A{t}}&=\mu_t\,,  & X_{ p {t}}&=0\,, \\ 
 X_{A}&= -\frac{1}{2 \alpha} \left( p+\beta \ast A \right) +  \mathrm{d}  A_{t}  \,,\qquad  & X_{p} &=  2 \alpha \delta \mathrm{d} A -  \frac{\beta }{2 \alpha} \ast \left( p+ \beta \ast A \right) -  \beta  \ast \mathrm{d}  A_{t} \,,
 \end{alignedat} \end{equation}
where the Dirac multiplier $\mu_t\in C^{\infty} (\Sigma)$ is arbitrary in the bulk.  This implies that $A_t$ is also arbitrary in the bulk, a fact which is, of course, related to the Abelian gauge symmetry
\begin{equation}
A \mapsto A+\d\epsilon\,, \quad p \mapsto p-\beta \ast \d\epsilon\,, \label{gautrans}
\end{equation}
with $\epsilon \in C^{\infty} (\Sigma)$ and $\imath_\partial^*(\epsilon)=0$ [these boundary conditions are a consequence of requiring $\imath_\partial^* \left(  A_{t}\right)=0$ in the definition of the configuration space].

\subsubsection*{Dirac analysis on the boundary}
The analysis of the boundary constraints strongly depends on $\lambda$; we show the final result below.

\paragraph{Case $\lambda=0$.}
After the first steps of the Dirac algorithm we obtain
\begin{subequations}
\begin{align}
 \imath_\partial^* \left(  A_{t}\right)=0\,, \label{eq:bcaz1}\\
\imath_\partial^*\left( \mu_t \right)=0\,, \label{eq:bcaz2} \\
\imath_\partial^* \left( \ast \mathrm{d} A \right)=0\,. \label{eq:bcaz3}
\end{align}
\end{subequations}
Remember that \eqref{eq:bcaz1} is a condition that was incorporated in the definition of the configuration space, the consistency condition derived from it gives \eqref{eq:bcaz2}, which fixes the value of $\mu_t$ at the boundary to zero. Equation \eqref{eq:bcaz3} is a secondary constraint at the boundary. Demanding its consistency we get the following infinite number of boundary constraints [see equation \eqref{infinite chain lambda 0} and its derivation in Appendix \ref{appendix infinite chain}]
\begin{subequations} \label{lzbc}
\begin{align}
       &\imath_\partial^* \left( \left( \ast \mathrm{d} \right)^{2k+1}   \left( p+\beta  \ast  A\right)    \right)=0\,, \label{eq:bcaz4}\\
&\imath_\partial^*    \left( \left(  \ast \mathrm{d} \right)^{2k+1}  A   \right)=0\,. \label{eq:bcaz8}
\end{align}
\end{subequations}

We pause now to make some comments. a) These kinds of constraints (an infinity chain of conditions) also appears in the case of a scalar field in manifolds with boundary \cite{Barbero_G_2016, Diracnos, Jabbari}. b) The actual number of boundary constraints in \eqref{lzbc} depends on the regularity demanded of the solutions to the field equations. As we are formally allowing for as much smoothness as we wish, we get an infinite tower of them.  c) Although similar conditions are introduced in the mathematical literature \cite{brezis2010} as necessary conditions to guarantee the smoothness of solutions to partial differential equations, usually they are not taken into account in the physical literature, in particular in the Hamiltonian analysis of field theories.

\paragraph{Case $\lambda\neq0$.}
After the first steps of the Dirac algorithm, we obtain
\begin{subequations}\allowdisplaybreaks
\begin{align}
\imath_\partial^* \left(  A_{t}\right)=0\,, \label{eq:bcanz1}\\
\imath_\partial^*\left( \mu_t \right)=0\,, \label{eq:bcanz2}\\
\imath_\partial^* \left( \{\alpha \ast \mathrm{d} +  \lambda^2 \iota_\nu \ast\} A \right)=0\,, \label{eq:bcanz3}\\
 \imath_\partial^*\Big(\{\alpha \ast \mathrm{d} + \lambda^2   \iota_\nu \ast\} \left(  p+\beta  \ast  A\right) \Big)=0\,. \label{eq:bcanz4}
\end{align}
\end{subequations}
The role of \eqref{eq:bcanz1}  and \eqref{eq:bcanz2} is the same as before. Equation \eqref{eq:bcanz3} is a secondary boundary constraint; its consistency gives rise to the new constraint \eqref{eq:bcanz4}. In this step, we have used $\imath_\partial^*\left(  \iota_\nu \ast \mathrm{d}  A_{t} \right)= - \ast_{\partial} \imath_\partial^*\left( \mathrm{d}  A_{t} \right)  =0$ which vanishes as a consequence of \eqref{eq:bcanz1}. Here and in the following $\ast_{\partial}$ denotes the Hodge dual in $\partial\Sigma$ with respect to the induced metric. As in the $\lambda=0$ case, the consistency of \eqref{eq:bcanz4} gives rise to an infinite chain of boundary constraints as explained in Appendix \ref{appendix infinite chain}.

For the particular case in which $\Sigma$ is a disk of radius $r_0$, redefining $\lambda$ so that $\lambda^2 \mapsto -  \alpha r_0 \lambda^2 $ with the new $\lambda$ a real constant, the constraint \eqref{eq:bcanz3} becomes
\begin{equation}
 \imath_\partial^* \left(  \ast \mathrm{d} A \right) = -\lambda^2  A_{\theta}\mid_{\partial}\,. \label{eq:bcbalach}
\end{equation}
This condition was introduced in Ref. \cite{Bala1994} after completing the Hamiltonian analysis of the action \eqref{MCS3}. This is why the constraints \eqref{eq:bcanz4} and the corresponding infinite chain were not considered there. We must say that, according to \cite{Bala1994}, if the disk is surrounded by a superconductor, then $1/\lambda^2$ can be interpreted as the penetration depth. This physical interpretation makes this model interesting, and for this reason, we will discuss it below.

We remark that we were able to obtain \eqref{eq:bcanz3} [or \eqref{eq:bcbalach}] as a natural boundary condition thanks to the boundary term that we included in the Lagrangian \eqref{eq:LagraMCSB}.

\subsection{Classical edge observables}\label{subsecceo}

Given $\Lambda \in C^{\infty} (\Sigma)$, we define the functions
\begin{equation}\label{eq:obsMCS}
Q_{\Lambda}(A,p)= \int_{\Sigma} \d \Lambda \wedge  \ast\left(  p- \beta \ast A  \right)\,.
\end{equation}
Under the gauge transformations of the theory \eqref{gautrans}, $A'= A+\d \epsilon, p'= p-\beta \ast \d\epsilon$, we have
\begin{equation*}
Q_{\Lambda}(A',p')= \int_{\Sigma} \d \Lambda \wedge  \ast \left(  p- \beta \ast A   - 2\beta \ast \d \epsilon  \right) = Q_{\Lambda}(A,p)-2\beta \int_{\partial \Sigma} \imath_\partial^*\left( \epsilon  \d \Lambda \right) = Q_{\Lambda}(A,p) \,.
\end{equation*}
Notice that the boundary term vanishes because $\imath_\partial^*\left( \epsilon\right)=0$ [which is a consequence of having incorporated the condition $\imath_\partial^*\left( A_t \right)=0$ in the definition of the configuration space]. We then conclude that the functions \eqref{eq:obsMCS} are observables characterized by the value of $\imath_\partial^*\left(\Lambda \right)$; because on the constraint submanifold defined by the condition $\delta \left( p-\beta \ast A \right)=0$ they can be written as a boundary integral
\begin{align*}
Q_{\Lambda}(A,p)&= \int_{\Sigma} \d \Lambda \wedge \ast\left(  p-\beta \ast A  \right) =   \int_{\Sigma}  \Lambda \ast \delta \left( p-\beta \ast A \right) +\int_{\partial \Sigma}  \imath_\partial^*\left( \Lambda  \ast \left(  p -\beta \ast A  \right)\right) \nonumber \\
&= \int_{\partial \Sigma}   \imath_\partial^*\left( \Lambda \ast \left( p -\beta \ast A  \right)\right) \,.
\end{align*}

We remark that the boundary conditions also play a role in the definition of the observables because these have to be evaluated on solutions to the Hamilton equations, which depend on them. 

With the help of the Hamiltonian vector field \eqref{VF33}, we get the evolution of the edge functions \eqref{eq:obsMCS}
\begin{equation}
\dot{Q}_{\Lambda}(A,p)= \int_{\Sigma} \d \Lambda \wedge  \ast\left(  X_p- \beta \ast X_A  \right)= 2\alpha \int_{\partial \Sigma}   \imath_\partial^*\left( \Lambda \d \ast \d A   \right) \,,
\end{equation}
where we have used $\imath_\partial^*\left(A_{t}\right)=0$. Notice that the edge functions \eqref{eq:obsMCS} are preserved in time for any $\Lambda$ if and only if  $ \d \imath_\partial^*\left( \ast \d A   \right) =0$. 
In the case $\lambda=0$, the primary boundary condition \eqref{eq:bcanz3} is $\imath_\partial^*\left( \ast \d A   \right)=0$. Therefore, for $\lambda=0$ the edge observables \eqref{eq:obsMCS} \textit{are constants of motion}. We will return to these observables after obtaining the solutions to the field equations in the next subsection.

Finally, the Poisson bracket of the two edge observables $Q_{\Lambda_1}(A,p)$ and $Q_{\Lambda_2}(A,p)$  is
\begin{align}
    \{ Q_{\Lambda_1}(A,p), Q_{\Lambda_2}(A,p)\}&= 2\beta \int_\Sigma  \d \Lambda_1 \wedge \d \Lambda_2= \beta \int_{\partial \Sigma}  \imath_\partial^* \left( \Lambda_1 \d \Lambda_2- \Lambda_2 \d \Lambda_1 \right)\,. \label{KacMoody}
\end{align}
Notice that, for $\beta=0$ they commute, but for $\beta\neq0$ and $\partial\Sigma \cong \mathbb{S}^1$ these observables generate the same $U(1)$ Kac-Moody algebra described in \eqref{KacMoodyCS}. Notice that, in the MCS theory, all the constraints are first class and hence \eqref{KacMoody} is the appropriate algebra. 

\subsection{Solving the Hamilton equations}

In this section, we determine the space of solutions to the Hamilton equations of motion in the phase space for $\lambda=0$ and discuss the peculiarities of the $\lambda\neq0$ case.  From now on we take $\alpha=-1/2$ (as is customary in the literature). In the bulk, the field equations in Hamiltonian form are
\begin{subequations}\label{eq: Hamiltonian}
\begin{align}
 \dot{A}_{t}&=\mu_t\,, \label{Hemblc1}\\  
 \dot{p}_{t}&=  0\,, \label{Hemblc2}\\
 \dot{A}&=  p+\beta \ast A  +  \mathrm{d}  A_{t}  \,, \label{Hemblc3} \\
 \dot{p} &=   - \delta \mathrm{d} A + \beta \ast  p- \beta^2  A  -  \beta  \ast \mathrm{d}  A_{t} \,, \label{Hemblc4}
\end{align}
\end{subequations}
with $\mu_t$ arbitrary.  Equation \eqref{Hemblc1} tells us that $A_t$ is arbitrary and \eqref{Hemblc2} tells us that $p_t$ is a constant of motion, which is actually zero because of the bulk constraint $p_t=0$. We remark that the fields $(A, p)$ must satisfy the bulk constraint 
\begin{equation}
\delta \left( p-\beta \ast A \right)=0\,. \label{eq:bcimp}
\end{equation}
In order to solve equations \eqref{Hemblc3}-\eqref{Hemblc4}, the constraint \eqref{eq:bcimp}, and the boundary constraints, our main tool will be the Hodge-Morrey theorem for manifolds with boundary \cite{Conner1956, abraham1993, schwarz2006}. This theorem will provide us with field decompositions that are specially appropriate for the problem that we are discussing here.

Let us introduce some definitions. We say that a form $\alpha$ is \textit{normal} if it has a vanishing tangential component, i.e., $\imath_\partial^* \alpha=0$, and \textit{tangential} if it has a vanishing normal component, i.e., $\imath_\partial^* \left( \ast \alpha \right) =0$.

\subsubsection[Case \texorpdfstring{$\lambda=0$}{lambda=0}]{Case \texorpdfstring{$\boldsymbol{\lambda=0}$}{lambda=0}}
A convenient decomposition for $\Omega^k(\Sigma)$ is given by the Hodge-Morrey theorem 
\[\Omega^k (\Sigma)= \mathcal{E}^k(\Sigma)\oplus \mathcal{C}^k (\Sigma)\oplus \mathcal{H}^k (\Sigma),\] 
where
\begin{subequations}\label{spacesb}
\begin{align}
   \mathcal{E}^k(\Sigma)&= \{  \mathrm{d} \gamma \, | \, \gamma \in \Omega^{k-1}(\Sigma) \,\, \textrm{with} \,    \imath_\partial^* \gamma=0 \} \,,\\
   \mathcal{C}^k (\Sigma)&= \{ \delta \zeta \, | \,  \zeta \in \Omega^{k+1}(\Sigma)  \,\, \textrm{with} \, \imath_\partial^* \left( \ast \zeta \right) =0 \} \,, \\
   \mathcal{H}^k (\Sigma)&= \{ h  \in \Omega^{k}(\Sigma) \, | \, \mathrm{d} h =0  \,\, \textrm{and} \,  \delta h =0\} \,.
\end{align}
\end{subequations}

Notice that, on $\Omega^1(\Sigma)$, $\ast$ satisfies $\ast^2=-\mathrm{id}$ so it endows $\Omega^1 (\Sigma)$ with the structure of a complex vector space that we denote as $\Omega^1 (\Sigma)_{\ast}$. The subspaces $\mathcal{E}^1(\Sigma)\oplus \mathcal{C}^1 (\Sigma)$ and $\mathcal{H}^1 (\Sigma)$ are complex subspaces of $\Omega^1 (\Sigma)_{\ast}$, i.e.,   $\ast ( \mathcal{E}^1(\Sigma)\oplus \mathcal{C}^1 (\Sigma))=\mathcal{E}^1(\Sigma)\oplus \mathcal{C}^1 (\Sigma)$ and $\ast \mathcal{H}^1 (\Sigma)=\mathcal{H}^1 (\Sigma)$. Finally,    $\ast\mathcal{E}^1(\Sigma)=\mathcal{C}^1(\Sigma)$ and $\ast \mathcal{C}^1(\Sigma)=\mathcal{E}^1(\Sigma)$. 

In the following, given a $k$-form $\eta$ we will write it as the sum $\eta=\eta_{\mathrm{d}}+ \eta_{\delta}+\eta_h$, with $\eta_{\mathrm{d}} \in \mathcal{E}^k (\Sigma),  \eta_{\delta}\in  \mathcal{C}^k (\Sigma), \eta_\mathrm{h} \in  \mathcal{H}^k (\Sigma)$. Then, for the 1-forms $A$ and $p$ we have
\begin{align} \label{hdap}
A= A_{\mathrm{d}}+ A_{\delta}+A_\mathrm{h}\,, \qquad p= p_{\mathrm{d}}+ p_{\delta}+p_\mathrm{h}\,,
\end{align}
Substituting \eqref{hdap} in the bulk constraint \eqref{eq:bcimp} gives  $\delta \left( p_{\mathrm{d}}-\beta \ast  A_{\delta} \right)=0$. Notice that we also have $\left( p_{\mathrm{d}}-\beta \ast A_{\delta} \right) \in \mathcal{E}^1 (\Sigma)$ [since $\ast \mathcal{C}^1(\Sigma)=\mathcal{E}^1(\Sigma)$]. In particular,  $\mathrm{d} \left( p_{\mathrm{d}}-\beta \ast A_{\delta} \right)=0$, then $p_{\mathrm{d}}-\beta \ast A_{\delta} \in  \mathcal{H}^1 (\Sigma)$, but $\mathcal{H}^1 (\Sigma) \cap \mathcal{E}^1 (\Sigma)= \{0\}$. Therefore, the bulk constraint \eqref{eq:bcimp} implies  
\begin{equation}
p_{\mathrm{d}}= \beta  \ast A_{\delta}\,. \label{cbcb}
\end{equation}

Before introducing the decomposition \eqref{hdap} into the Hamilton equations, notice that in this case the first boundary constraint \eqref{eq:bcaz3} is 
\begin{equation}
0=\imath_\partial^* \left( \ast \mathrm{d} A \right)=\imath_\partial^* \left( \ast \mathrm{d} A_\delta \right)\,.\label{bcaz67}
\end{equation}
Then we have that $\delta \mathrm{d} A=\delta \mathrm{d} A_\delta\in  \mathcal{C}^1 (\Sigma)$. Actually, it is straightforward to prove the converse: $\left(\delta \mathrm{d} A \right)_{\textrm{d}}=0$ and  $\left(\delta \mathrm{d} A \right)_{\textrm{h}}=0$ implies $\imath_\partial^* \left( \ast \mathrm{d} A \right)=0$. 

Using \eqref{hdap}, \eqref{cbcb}, and $\delta \mathrm{d} A_\delta\in  \mathcal{C}^1 (\Sigma)$ allows us to write the independent set of equations of motion \eqref{Hemblc3}-\eqref{Hemblc4} as
\begin{subequations}\allowdisplaybreaks
\begin{align}
 \ddot{A}_{\delta}  &=  - \left(\delta \mathrm{d} + 4 \beta^2 \right) A_\delta\,,  \label{eq:hdeq2}\\
\dot{A}_{\mathrm{d}} &=   2\beta  \ast A_{\delta} +  \mathrm{d}  A_{t} \,,  \label{eq:hdeq1}\\
     p_\delta &=  \dot{A}_{\delta} - \beta \ast A_{\mathrm{d}} \,, \label{eq:hdeq3}\\
     \dot{A}_\mathrm{h}&= p_\mathrm{h} + \beta \ast A_\mathrm{h}\,, \label{eq:hdeq4}\\
    \dot{p}_\mathrm{h}&= \beta \ast p_\mathrm{h} - \beta^2 A_\mathrm{h}\,.\label{eq:hdeq5}
\end{align}
\end{subequations}

Notice that the components $A_\mathrm{h}$ and $A_\delta$ are decoupled (we show below that they parametrize the reduced phase space) and that if we find $A_\delta$, then we can directly calculate $A_{\mathrm{d}}$, $p_{\mathrm{d}}$, and $p_\delta$. This suggests that, in order to solve \eqref{eq:hdeq2} together with the boundary constraint \eqref{bcaz67}, we should first look for $\vartheta \in \Omega^1(\Sigma)$ satisfying 
\begin{equation}\label{varphinb1}
\delta \mathrm{d} \vartheta=  \omega^2 \vartheta  \quad \textrm{with}  \quad \imath_\partial^*\left(   \ast \mathrm{d}   \vartheta \right)=0\,.
\end{equation}
This is a well-posed problem in the sense that, under these conditions, the positive-definite operator $\delta \mathrm{d}$ is self-adjoint \cite{Conner1956}. Hence, according to the spectral theorem, there always exists an orthonormal basis of eigen 1-forms $\vartheta$. Notice that, when $\omega\neq0$, equation  \eqref{varphinb1} implies $\vartheta\in\mathcal{C}^1(\Sigma)$ and for $\omega=0$ we have $\vartheta\in\mathcal{E}^1(\Sigma)\oplus \mathcal{H}^1 (\Sigma)$. This is so because  $\delta \mathrm{d} \vartheta= 0$ implies that $\ast \mathrm{d}   \vartheta$ is a constant (which is actually zero as a consequence of the boundary condition), hence, $\mathrm{d}\vartheta=0$.

For $\omega \neq 0$, using the eigen 1-forms defined in \eqref{varphinb1} it is possible to find the solutions $A_\delta$ of \eqref{eq:hdeq2}. However, before we write them, it helps to cast \eqref{varphinb1} in a more familiar form. Let us define the function $F:=  \ast \mathrm{d}   \vartheta$.
Taking into account that $\vartheta$ is an eigen 1-form, $\delta \mathrm{d} \vartheta=  \omega^2 \vartheta$, we have
\begin{equation}\label{varphiandF}
     \ast \mathrm{d} F= - \omega^2 \vartheta \Rightarrow  \vartheta= -\frac{\ast \mathrm{d} F }{ \omega^2}     \,;
\end{equation}
Therefore, we only have to find the function $F$ to determine $\vartheta$.  Using  \eqref{varphiandF}, the conditions \eqref{varphinb1} are equivalent to
\begin{equation}
\nabla^2 F =-\omega^2 F  \quad \textrm{with}  \quad \imath_\partial^*\left( F \right)=0 \,,\label{eqfF2}
\end{equation}
where, acting on functions, $\delta \mathrm{d}=-\nabla^2$ is minus the standard (nonpositive) scalar Laplacian. In order to give an explicit solution of \eqref{eqfF2} we need to specify $\Sigma$. Notice that $\vartheta$ must be of the form $\delta \phi$ with $\imath_\partial^*\left( \ast \phi  \right)=0$. From \eqref{varphiandF}, we can write $\phi=    \ast F / \omega^2$, which already satisfies $\imath_\partial^*\left( \ast \phi  \right)=0$ as a consequence of  $\imath_\partial^*\left(  F \right)=0$.

Let us assume that the eigen 1-forms in \eqref{varphinb1} exist and denote them as $\vartheta_I$ (with eigenvalue $\omega_I^2>0$). Using this orthonormal basis, $\langle \vartheta_I, \vartheta_J \rangle= \int_{\Sigma} \vartheta_I \wedge \ast \vartheta_J=\delta_{IJ}$, the solutions to \eqref{eq:hdeq2}-\eqref{eq:hdeq3} are
\begin{subequations} \allowdisplaybreaks\label{solexp2}
\begin{align}
A_\delta(t)=& \sum_{I}  \frac{1}{\sqrt{2 \Tilde{\omega}_{I}}} \big( C_{I} \exp{(i \Tilde{\omega}_{I} t)} +  C^*_{I} \exp{(-i \Tilde{\omega}_{I} t)} \big)  \vartheta_{I} \,,\\
p_\delta(t) =& \sum_{I}  \frac{i }{\sqrt{2 \Tilde{\omega}^3_{I}}}  \Big( \left(  \Tilde{\omega}^2_{I}-2\beta^2 \right) \big(  C_{I} \exp{(i \Tilde{\omega}_{I} t)} -  C^*_{I} \exp{(-i \Tilde{\omega}_{I} t)}\big)+ 2 \beta^2 \left( C_{I}-C^*_{I} \right)\Big) \vartheta_{I} \nonumber\\
 &   - \beta \ast  \left( \mathrm{d} \left( \int_0^t A_t \mathrm{d}t' \right) +A_{\mathrm{d}}(0)\right)\,,\\
A_{\mathrm{d}}(t) =& 2 \beta \sum_{I}  \frac{-i}{\sqrt{2 \Tilde{\omega}^3_{I}}} \big( C_{I} \left( \exp{(i \Tilde{\omega}_{I} t)} -1\right)    - C^*_{I} \left( \exp{(-i \Tilde{\omega}_{I} t)}-1\right) \big)   \ast  \vartheta_{I} \nonumber\\
&+ \mathrm{d} \left( \int_0^t A_t \mathrm{d}t' \right)+A_{\mathrm{d}}(0)\,,\\
p_{\mathrm{d}}(t)=&  \beta \sum_{I}   \frac{1}{\sqrt{2 \Tilde{\omega}_{I}}} \big( C_{I} \exp{(i \Tilde{\omega}_{I} t)} +  C^*_{I} \exp{(-i \Tilde{\omega}_{I} t)}\big)  \ast  \vartheta_{I} \,,
\end{align}
\end{subequations}
with $\Tilde{\omega}_{I}^2= \omega^2_{I} + 4 \beta^2$. The real and imaginary parts of the complex constant $C$ are given by
\begin{equation*}
\sqrt{\frac{2}{\Tilde{\omega}_{I}}} \, \textrm{Re} \,C_{I}= \langle  \vartheta_{I}, A_\delta(0) \rangle\,, \quad
- \sqrt{2 \Tilde{\omega}_{I}} \,\textrm{Im}\, C_{I}  =  \langle \vartheta_{I}, p_\delta(0)+ \beta  \ast A_d (0) \rangle \,.
\end{equation*}

Notice that, so far, we have only used the boundary constraint \eqref{eq:bcaz3}, not the infinite chain \eqref{lzbc}. However, it must be remarked that \textit{all} the constraints in \eqref{lzbc} are satisfied if $\imath_\partial^*\left(   \ast \mathrm{d}   \vartheta_I \right)=0$. Actually, plugging \eqref{hdap} and \eqref{solexp2} in \eqref{eq:bcaz4} and \eqref{eq:bcaz8}, we get
\begin{align*}\allowdisplaybreaks
     \imath_\partial^* \left( \left( \ast \mathrm{d} \right)^{2n+1}   \left( p_\delta +\beta  \ast  A_{\textrm{d}}\right)    \right) 
    & \propto \sum_{I} (-1)^n \omega_I^{2n}  \imath_\partial^*\left( \ast \mathrm{d}  \vartheta_I  \right)  \,, \\
    \imath_\partial^*\left( \left(  \ast \mathrm{d} \right)^{2n+3}   A_\delta    \right)
    &\propto  \sum_{I} (-1)^{n+1}  \omega_I^{2n+2} \imath_\partial^* \left(   \ast \mathrm{d}   \vartheta_I\right)\,.
\end{align*}
Then, our solutions \eqref{solexp2} actually satisfy the infinite chain of boundary conditions \eqref{lzbc}. This situation is similar to the scalar field case \cite{Barbero_G_2014, Jabbari}.

We study now the harmonic sector. First, one should notice that the harmonic 1-forms satisfy all the boundary conditions \eqref{eq:bcaz3}-\eqref{lzbc}. The evolution equations are \eqref{eq:hdeq4} and \eqref{eq:hdeq5}. In order to solve them, we first notice that $\dot{p}_\mathrm{h}-\beta \ast \dot{A}_\mathrm{h}=0$, hence the $p_\mathrm{h}-\beta \ast A_\mathrm{h}$ are constants of motion. They will play a relevant role in the edge observables discussed below. Second, we define $\pi_h:= p_\mathrm{h} + \beta \ast A_\mathrm{h}$, then  \eqref{eq:hdeq4} and \eqref{eq:hdeq5} are equivalent to
\begin{equation*}
    \dot{A}_\mathrm{h}= \pi_\mathrm{h} \,,  \qquad    \dot{\pi}_\mathrm{h}= 2 \beta \ast \pi_\mathrm{h} \Rightarrow \ddot{\pi}_\mathrm{h}= -4\beta^2 \pi_\mathrm{h}  \,,
\end{equation*}
whose solutions are
\begin{align*}
   \pi_\mathrm{h}(t)&= \left( \cos\left( 2\beta t \right)+ \sin\left( 2\beta t \right) \ast \right) \pi_\mathrm{h}(0)\,,\\
   A_\mathrm{h} (t)&= A_\mathrm{h} (0) +  \frac{1}{2\beta}  \left( \sin\left( 2\beta t \right)+ \left(1- \cos\left( 2\beta t \right) \right) \ast \right) \pi_\mathrm{h}(0) \,.
\end{align*}
In terms of $A_\mathrm{h}, p_\mathrm{h}$ we get
\begin{subequations}\label{solexp22}
\begin{align}
   A_\mathrm{h} (t)&\!=\!  \frac{1}{2\beta} \!\ast \Big( p_\mathrm{h}(0)- \beta \ast\! A_\mathrm{h} (0)  \Big)  +  \frac{1}{2\beta}  \Big( \sin\left( 2\beta t \right)- \cos\left( 2\beta t \right) \!\ast\! \Big) \Big(p_\mathrm{h}(0)+ \beta \!\ast\! A_\mathrm{h} (0)  \Big)\,, \\
    p_\mathrm{h}(t)&\!=\!\frac{1}{2}  \Big( p_\mathrm{h}(0)- \beta \ast A_\mathrm{h} (0)  \Big)+\frac{1}{2} \Big( \cos\left( 2\beta t \right) +\sin\left( 2\beta t \right) \ast \Big) \Big(p_\mathrm{h}(0) + \beta \ast A_\mathrm{h} (0)  \Big)\,.
\end{align}
\end{subequations}
As mentioned before the 1-forms
\[
p_\mathrm{h}(t)-\beta \ast  A_\mathrm{h} (t)= p_\mathrm{h}(0)-\beta \ast  A_\mathrm{h} (0)
\]
are time independent. 

Notice that as $\ast$ is a (linear) complex structure on $\mathcal{H}^1(\Sigma)$, there exists a complex infinite (but countable) orthonormal basis $\{h_m, \bar{h}_m\}$ ($m \in \mathbb{N}$ and the bar over the 1-forms $h_m$ denotes their complex conjugate) of $\mathcal{H}^1(\Sigma)$ formed by the eigen 1-forms of $\ast$ (as $\ast^2= -1$ acting over 1-forms, the eigenvalues are $\pm i$), i.e., the $h_m$ satisfy
\begin{equation}
\ast h_m=  - i h_m\,, \quad   \ast \bar{h}_m=  i \bar{h}_m \,, \quad (  h_m ,  h_l ) = \delta_{m l} =  (  \bar{h}_m ,  \bar{h}_l ) \,, \quad (  h_m ,  \bar{h}_l )=0\,, \label{hamocondib}
\end{equation}
where $(  h_m ,  h_l ) = \int_{\Sigma} \bar{h}_m \wedge \ast h_l$. Using this basis, for $\beta>0$ we can write
\begin{align*}
    p_\mathrm{h}(0)+\beta \ast  A_\mathrm{h} (0)&= \sqrt{2 \beta}  \sum_m \left(  a_m h_m+  a^*_m \bar{h}_m\right)\,, \\ p_\mathrm{h}(0)-\beta \ast  A_\mathrm{h} (0)&= \sqrt{2 \beta}  \sum_m \left(  b^*_m h_m+  b_m \bar{h}_m\right)\,,
\end{align*}
with
\begin{align}
    a_m&=  \frac{1}{\sqrt{2 \beta}} (h_m,  p_\mathrm{h}(0)+\beta \ast  A_\mathrm{h} (0)) \,,&& a^*_m= \frac{1}{\sqrt{2 \beta}} (\bar{h}_m,  p_\mathrm{h}(0)+\beta \ast  A_\mathrm{h} (0))\,, \nonumber\\
   b_m&= \frac{1}{\sqrt{2 \beta}} (\bar{h}_m,  p_\mathrm{h}(0)-\beta \ast  A_\mathrm{h} (0))\,,&& b^*_m=  \frac{1}{\sqrt{2 \beta}} (h_m,  p_\mathrm{h}(0)-\beta \ast  A_\mathrm{h} (0))\,,
\end{align}
while for $\beta<0$ we must change $\sqrt{2 \beta} \to  \sqrt{-2 \beta}$, and interchange $a_m$ with $a^*_m$ and $b_m$ with $b^*_m$.  
This allows us to write $A_\mathrm{h} (t)$ and $p_\mathrm{h} (t)$ in terms of this basis.

\paragraph{The disk}

In order to give explicit expressions of the eigen 1-forms $\vartheta$ in \eqref{varphinb1}, we consider the case in which $\Sigma$ is a disk of radius $r_0$. Using polar coordinates $(r, \theta)$ and separation of variables, we write $F(r,\theta)= g(\theta) f(r)$ [with $g(0)=g(2\pi)$] in \eqref{eqfF2} to get
\begin{subequations} \label{sepavar}
\begin{align}
g''(\theta)&=-N^2 g(\theta) \,, \label{sepavari1} \\
\left( \frac{\partial^2}{\partial r^2}+ \frac{1}{r} \frac{\partial}{\partial r}+\left(  \omega^2- \frac{N^2}{r^2}\right) \right) f (r)&=0 \,. \label{sepavari2}
\end{align}
\end{subequations}
where $N$ is a constant. The solutions to \eqref{sepavari1} are of the form $\exp{\left( i N \theta \right)}$ with $N \in \mathbb{Z}$. Equation \eqref{sepavari2} is the Bessel equation; its finite solutions at $r=0$ are the $J_N(\omega r)$ Bessel's functions.  Thus, the solutions $F(r,\theta)$ to \eqref{eqfF2} can be written in terms of the $\exp{\left( i N \theta \right)}  J_N(\omega r)$, which must satisfy the boundary condition $\imath_\partial^*\left( F \right)=0$. This implies $J_N(\omega r_0)=0$, which tells us what the values of $\omega$ are. As we can see, for each $N$ we have a family of $\omega$'s.  We denote these infinite (but countable) sets by $\omega_{N, n}$ (equal to $z_{N,n}/r_0$, where $z_{N,n}$ are the zeros of $J_N$). The index $I$ used in the previous subsection corresponds now to the pair $(N, n)$. 


We conclude that the real eigen 1-forms $\vartheta_{N, n}$ are
\begin{equation*}
   \vartheta_{N, n}= \frac{1}{\omega^2_{N,n}} \ast \mathrm{d} \Big(  \left(  A_{N, n} \exp{\left( i N \theta \right)}   + A^*_{N, n} \exp{\left( -i N \theta \right)} \right) J_N(\omega_{N, n} r)\Big)\,.
\end{equation*}
The complex constants $A_{N, n}$ are fixed by the orthonormality condition $\langle \vartheta_{N,n}, \vartheta_{M,m} \rangle= \delta_{n m} \delta_{N M}$ (the $\delta_{N M}$ is a consequence of Bourget's hypothesis, a corollary of a theorem proved by Carl Ludwig Siegel \cite{watson}). Notice that we must replace $\sum_I$ by $\sum_{N} \sum_{n}$ in the solutions \eqref{solexp2}.

Finally, in this case, the harmonic forms $h_n$ satisfying \eqref{hamocondib} are  
\begin{equation}
   h_n=\frac{1}{\sqrt{2 \pi n} r_0^n}  \mathrm{d} z^n \,,\label{hn}
\end{equation}
with $z= x_1+i x_2$  (here $x_1,x_2$ are Cartesian coordinates in $\Sigma$) and $n \in \mathbb{N}$ \cite{Bala1994}. Notice that, using polar coordinates,
\[
h_n=\sqrt{\frac{n}{2\pi}}\left(\frac{r}{r_0} e^{i\theta}\right)^n\left(\frac{\mathrm{d} r}{r}+i \mathrm{d}\theta\right)\,,
\]
and it is straightforward to check that, for $r<r_0$, $h_n \xrightarrow[n\rightarrow\infty]{} 0$. On the other hand, for $r=r_0$ we get $h_n=\sqrt{\frac{n}{2\pi}}\ e^{n\theta}\left(\frac{\mathrm{d} r}{r_0}+i \mathrm{d}\theta\right)$. Hence the eigen 1-forms $h_n$ behave as classical edge states in the sense of references \cite{Asorey2013,Asorey2016}

\subsubsection[Case \texorpdfstring{$\lambda\neq0$}{lambda not zero}]{Case \texorpdfstring{$\boldsymbol{\lambda\neq0}$}{lambda not zero}}

Regardless of the boundary conditions, we have shown that the decomposition \eqref{hdap} can be used to solve the bulk constraint in a convenient way [obtaining \eqref{cbcb}]. However, for $\lambda\neq0$, the boundary condition \eqref{eq:bcanz3} is
\begin{equation}\label{bcl1}
\imath_\partial^* \left( \{ \ast \mathrm{d}  -  2  \lambda^2 \iota_\nu \ast\} A \right)=0\,,
\end{equation}
which is different from the one that appears in the previous case where we had $\imath_\partial^* \left(  \ast \mathrm{d} A\right)=0$. As a consequence, we have now $\left(\delta \mathrm{d} A \right)_{\textrm{h}}\neq 0$, thus $\delta \mathrm{d} A = \left(\delta \mathrm{d} A \right)_{\delta}+ \left(\delta \mathrm{d} A \right)_{\textrm{h}}$ with both components different from zero. Notice that using \eqref{hdap} we can write \eqref{bcl1} as
\begin{equation}\label{bclcomb}
\imath_\partial^*\left(  \ast \mathrm{d} \left(A_\delta+A_\mathrm{h}\right) \right) +  2  \lambda^2 \ast_{\partial}  \imath_\partial^*\left(A_\delta+A_\mathrm{h} \right)=0\,.
\end{equation}
This leads us to work with the combination $A_\delta+A_\mathrm{h}=: A_{\delta \mathrm{h}}$. Using the Hamiltonian equations \eqref{Hemblc3} and \eqref{Hemblc4} we see that $A_{\delta \mathrm{h}}$ must satisfy
\begin{equation}\label{eq:hdeq2comb}
    \ddot{A}_{\delta \mathrm{h}} =  - \left(\delta \mathrm{d} + 4 \beta^2 \right) A_{\delta \mathrm{h}}+ 2\beta \ast \left(p_\mathrm{h} - \beta \ast A_\mathrm{h}\right)\,.
\end{equation}
The presence of $p_\mathrm{h} - \beta \ast A_\mathrm{h}$ in \eqref{eq:hdeq2comb} makes it very difficult to solve because this term involves a projector onto the harmonic sector, which is related to a (nonlocal) Green's operator. We remark that in the $\lambda\neq0$ case the $p_\mathrm{h} - \beta \ast A_\mathrm{h}$ are no longer constants of motion because $\dot{p}_\mathrm{h} - \beta \ast \dot{A}_\mathrm{h}=- \left(\delta \mathrm{d} A_\delta \right)_{\textrm{h}}$. In the pure Maxwell case $\beta=0$, it is possible to use the eigen 1-forms of the operator $\delta \mathrm{d}$ to solve \eqref{eq:hdeq2comb}, i.e.,
\begin{equation} \label{eigprob2}
\delta \mathrm{d} \vartheta=  \omega^2 \vartheta \quad \textrm{with} \quad  \imath_\partial^*\left(  \ast \mathrm{d}  \vartheta  -  2  \lambda^2 \iota_\nu \ast \vartheta \right)=0\,.
\end{equation}
This is a well-posed problem and the operator $\delta \mathrm{d}$ with these Robin-like boundary conditions is self-adjoint \cite{Bala1994}. The corresponding spectrum and eigenfunctions when $\Sigma$ is a disk were (partially) analyzed in \cite{Bala1994}. This is an interesting problem by itself. In Appendix \ref{apenA}, we show how to deal with \eqref{eigprob2} from the Hodge decomposition point of view.

Another strategy to solve the Hamiltonian equations is to use a different Hodge-like decomposition adapted to the boundary constraints \eqref{bcl1}. For instance, we can write $A$ and $p$ as 
\begin{equation}
A=  A_{\delta}+A_\textrm{cn} \,, \qquad p=  p_{\delta}+p_\textrm{cn} \,,\label{hdap2}
\end{equation}
where $A_{\delta}=\delta \phi$, [with $\phi \in \Omega^{2}(\Sigma)$ and free at the boundary $\partial \Sigma$], and $A_\textrm{cn}$ and $p_\textrm{cn}$ closed 1-forms normal to $\partial \Sigma$, (i.e., $\mathrm{d} A_{cn}=0=\mathrm{d} p_{cn}$ and  $\imath_\partial^* A_\textrm{cn}=0=\imath_\partial^* p_\textrm{cn}$)  \cite{Conner1956, abraham1993}. Notice that as $A_\textrm{cn}$ is normal to $\partial \Sigma$  then $\imath_\partial^* \left( \iota_\nu \ast  A_\textrm{cn}\right)= \ast_{\partial}  \imath_\partial^* A_\textrm{cn} =0$. Taking all this into account and using the decomposition \eqref{hdap2} on the boundary constraint \eqref{eq:bcanz3} we get
\begin{align*} 
 \imath_\partial^*\left(  \ast \mathrm{d}  A_{\delta}  -  2  \lambda^2 \iota_\nu \ast A_{\delta} \right)=0 \,.
\end{align*}
This equation only involves $A_{\delta}$, which seems to suggest the use of the composition \eqref{hdap2} to solve the Hamilton equations of motion. 

Plugging \eqref{hdap2} into the constraint $\delta \left( p-\beta \ast A \right)=0$, we obtain $\d \ast \left( p_\textrm{cn}-\beta \ast A_\delta \right)=0$. Assuming that $\Sigma$ is a smoothly contractible manifold with boundary (then, its first de Rham cohomology group is zero  \cite{WEINTRAUB2014361}), the previous equation implies  $\left( p_\textrm{cn}-\beta \ast A_\delta \right)_{\textrm{cn}}=0$, and we get $p_\textrm{cn}=\beta \left( \ast A_{\delta}\right)_{\textrm{cn}}$. Unfortunately,  $\left( \ast A_{\delta}\right)_{\textrm{cn}} \neq \ast A_{\delta}$, and then the solution to the bulk constraint involves a nonlocal operator (the projector onto the space of the closed 1-forms normal to $\partial \Sigma$). Once again, the problem becomes intractable.

\subsection{Fock quantization and quantum edge observables} \label{sec:4}

In this section we present the (\textit{reduced phase space}) Fock quantization \cite{juarez2015quantization} of the MCS model for the case $\lambda=0$ and study the corresponding quantum edge  observables. We start by computing the pullback of the symplectic structure $\Omega=   \int_\Sigma  \dd A \www \ast \dd p$ to the space of solutions given by \eqref{solexp2} and \eqref{solexp22}, which we denoted $\Omega_S$. The result is
\begin{align} 
 \Omega_S&=   \int_\Sigma \left( \dd A_\mathrm{d} \www \ast \dd p_\mathrm{d} + \dd A_\delta \www \ast \dd p_\delta + \dd A_\mathrm{h} \www  \ast \dd  p_\mathrm{h}  \right) \nonumber\\
   &= -i\sum_{I}    \dd C_I \www \dd C^*_I+ \int_\Sigma  \dd  A_\mathrm{h} (0) \www  \ast    \dd p_\mathrm{h} (0) \nonumber\\
    &= -i\sum_{I}    \dd C_I \www \dd C^*_I+ \sum_m  \left( -i \dd a_m \www \dd a^*_m - i \dd b_m \www \dd b^*_m \right)\,.
\end{align}
The pullback of the Hamiltonian \eqref{Hamiltonian_MCSB} to this space in the $\lambda=0$ case is 
\begin{align}
    H_S&= \sum_{I}  \Tilde{\omega}_I C^*_I C_I  + \frac{1}{2} \int_\Sigma  \left( p_\mathrm{h}(0)+\beta \ast A_\mathrm{h} (0) \right) \wedge  \ast   \left( p_\mathrm{h} (0)+\beta \ast A_\mathrm{h} (0) \right) \nonumber \\
    &=\sum_{I}  \Tilde{\omega}_I C^*_I C_I  + 2 \abs{\beta} \sum_m a^*_m a_m \,. \label{reduHa}
\end{align}
Therefore, as it must be clear from the previous expressions, we end up with an infinite number of uncoupled harmonic oscillators, one harmonic oscillator of frequency $\Tilde{\omega}_I$ for each eigen 1-form $\vartheta_I$ and an infinite number of oscillators of frequency $2\abs{\beta}$ in the harmonic sectors. Notice that the $b_m$-modes are constants of motion. 

The Fock quantization of the system is direct: we promote the variables $C_I$, $C^*_I$, $a_m$, $a^*_m$, $b_m$, and $b^*_m$ to creation and annihilation operators $\hat{C}^{\dag}_I$, $\hat{a}^{\dag}_m$, $\hat{b}^{\dag}_m$ and $\hat{C}_I$, $\hat{a}_m$, $\hat{b}_m$, respectively, with nonvanishing commutators given by
\begin{align*}
   [ \hat{C}_I  , \hat{C}^{\dag}_J]= \delta_{IJ}, \qquad [ \hat{a}_m  , \hat{a}^{\dag}_n]= \delta_{mn}= [ \hat{b}_m  , \hat{b}^{\dag}_n]\,.
\end{align*}
By using these operators, we get  from \eqref{reduHa} the quantum Hamiltonian operator
\begin{equation}
    \hat{H}    =\sum_{I}  \Tilde{\omega}_I  \hat{C}^{\dag}_I \hat{C}_I + 2\abs{\beta} \sum_n \hat{a}^{\dag}_n \hat{a}_n\,. 
\end{equation}

The basis states for the theory are taken as the product of the basis states 
for each oscillator (which can be chosen to be the eigen states of the number operators $\hat{C}^{\dag}_I \hat{C}_I$, $\hat{a}^{\dag}_n \hat{a}_n$, and  $\hat{b}^{\dag}_n \hat{b}_n $).

\paragraph{Quantum edge observables}  

The quantum edge observables are obtained by promoting, when possible, the classical edge observables to operators. In this case, the observables \eqref{eq:obsMCS} evaluated on the solutions \eqref{solexp2} and \eqref{solexp22} reduce to 
\begin{align}\label{eq:obsMCSoversol}
Q^S_{\Lambda}(A,p)&= \int_{\Sigma} \d \Lambda \wedge  \ast\Big( p(t)-\beta \ast A(t)  \Big)\nonumber\\
&=\int_{\partial \Sigma}   \imath_\partial^*\Big( \Lambda \ast \big(p_\delta(t)-\beta \ast A_{\mathrm{d}}(t)+ p_\mathrm{h}(0)- \beta \ast A_\mathrm{h}(0)  \big)\Big) \nonumber\\
&=\int_{\Sigma} \d \Lambda \wedge  \ast\Big( p_\mathrm{h}(0)- \beta \ast A_\mathrm{h}(0)  \Big)\,,
\end{align}
where we have used $p_{\mathrm{d}}= \beta  \ast A_{\delta}$, $\imath_\partial^*\left( \ast p_\delta \right)=0$, and $\imath_\partial^* \left(A_{\mathrm{d}} \right)=0$. In subsection \eqref{subsecceo}, we have shown that, for $\lambda=0$, the edge observables are constants of motion (which correspond to the constants of motion in the harmonic sector found in the previous section). In fact, by using the basis $\{ h_m, \bar{h}_m\}$, we can write \eqref{eq:obsMCSoversol} for $\beta>0$ as
\begin{align}
 Q^S_{\Lambda}(A,p)   &= \sqrt{2 \beta}   \sum_m  \int_{\Sigma} \d \Lambda \wedge \ast \big( b^*_m h_m +b_m \bar{h}_m  \big) \nonumber\\
 &= \sqrt{2 \beta}   \sum_m \left(  b^*_m  \int_{\Sigma} \d \Lambda \wedge \ast  h_m +b_m   \int_{\Sigma} \d \Lambda \wedge \ast \bar{h}_m  \right)\,, \label{eor}
\end{align}
and an analogous expression for $\beta<0$.

From \eqref{eor}, we define the quantum edge observable of the MCS theory acting over the harmonic basis   $\{ h_n, \bar{h}_n\}$ as
\begin{subequations}\label{qeor}
\begin{align}
\hat{Q}_{h_n} &:= \sqrt{2 \beta}   \sum_m \left(  \hat{b}^\dag_m  \int_{\Sigma} h_n \wedge \ast  h_m + \hat{b}_m   \int_{\Sigma} h_n \wedge \ast \bar{h}_m  \right)\nonumber\\
&=\sqrt{2 \beta}   \sum_m \left(  \hat{b}^\dag_m   (  \bar{h}_n ,  h_m )+ \hat{b}_m   (  \bar{h}_n ,  \bar{h}_m ) \right)\nonumber\\
&=\sqrt{2 \beta} \,   \hat{b}_n \,,\\
\hat{Q}_{\bar{h}_n} &:= \sqrt{2\beta}   \sum_m \left(  \hat{b}^\dag_m   (  h_n ,  h_m )+ \hat{b}_m   (  h_n ,  \bar{h}_m ) \right)\nonumber\\
&=\sqrt{2 \beta} \,  \hat{b}^\dag_n \,,
\end{align}
\end{subequations}
where we have used $ (  h_m ,  h_l ) = \delta_{m l} =  (  \bar{h}_m ,  \bar{h}_l )$, $(  h_m ,  \bar{h}_l )=0$. For $\beta<0$ we obtain  $\hat{Q}_{h_n}= \sqrt{-2\beta} \,   \hat{b}^\dag_n$ and $\hat{Q}_{\bar{h}_n}=\sqrt{-2\beta} \,   \hat{b}_n$. As pointed out in \cite{Bala1994}, the Fock states created by the operators $\hat{b}^\dag_n$ can be thought of as quantum states localized at the boundary. As expected, the quantum edge observables correspond to (linear combinations of) the operators $\hat{b}_n, \hat{b}^\dag_n$ which are obtained by promoting to quantum objects the $b_n, b^*_n$-modes (remember that these are constants of motion).

We end by pointing out that, for the particular case in which the fields are defined on a disk, the quantization in \textit{the full phase space} of the MCS action was discussed in \cite{Bala1994}. There, the authors followed the rules of Dirac's quantization, imposed the Gauss law as an operator that annihilates physical states, and tried to diagonalize it together with the Hamiltonian. They succeeded for $\lambda=0$, but not for $\lambda\neq 0$. Despite the differences in the in the approach of \cite{Bala1994} and ours, for the particular case of the disk and $\lambda=0$, the results about the classical \eqref{eor} and quantum \eqref{qeor} edge observables are the same.

\section{Conclusions}\label{sec:conclu}

We have used the Abelian Chern-Simons model to illustrate some classical aspects of the so-called edge observables. Then, we have studied in detail the Lagrangian and the Hamiltonian formulations of the Maxwell-Chern-Simons model defined on a manifold with boundary for two different sets of boundary equations naturally derived from a variational principle. Using the geometric version of the Dirac algorithm, we have been able to handle in a rigorous way the introduction of the boundary and obtain the infinite chain of boundary constraints of the model, which are usually ignored in the literature. 

We have shown that, inspired by the Gauss constraint (which is first class), one can build classical edge observables. Their formal definition is independent of the boundary conditions imposed on the field variables, but their actual values and properties depend on them. We have shown that for $\lambda=0$ these observables are constants of motion, while for $\lambda \neq 0$ they are not. Also, we have calculated their algebra which, when the boundary of $\Sigma$ is a circumference, is the well-known $U(1)$ Kac-Moody algebra.

In order to get a better characterization of the classical edge observables and states and perform the Fock quantization of the MCS model, we have looked for the solutions of the Hamilton equations of motion together with the bulk and boundary constraints. Our principal tool has been the Hodge-Morrey decomposition. For $\lambda=0$, we have found the solutions, without any gauge fixing, and showed their explicit form for the particular case in which the fields are defined on a disk. For the case $\lambda \neq 0$, we have discussed the obstructions that prevent us from obtaining the corresponding solutions by using the procedure that works in the $\lambda=0$ case.

For $\lambda=0$, we have shown that, on the space of solutions, the system reduces to an infinite collection of uncoupled oscillators. This has allowed us to directly carry out the Fock quantization. Furthermore, we have discussed the classical and quantum edge observables. In the reduced phase space, they correspond to the constants of motion of the harmonic sector and the quantum operators associated with these constants, respectively.  Explicitly, when $\Sigma$ is a disk, the Hodge dual endows the harmonic sector (of the Hodge-Morrey decomposition) with a basis of eigen 1-forms \eqref{hn} that can be identified with  the classical edge states (at least when $n$ is large). These states can be used not only to expand the constant of motion \eqref{eq:obsMCSoversol}   [see \eqref{eor}] but also to define a privileged pair of edge observables \eqref{qeor}.

Our results can be applied to other compact regions besides the disk, in particular the resolution of the field equations for $\lambda=0$. The case with noncompact $\Sigma$ is also interesting and has been considered in the literature   (see, for instance, \cite{Blasi2010}), but the spectra of some relevant operators become continuous and the analytical issues that crop up must be carefully considered. 

The strategy that we have followed in the present work can be used in principle for other boundary conditions for the MCS model. As far as the edge observables are concerned, it would be interesting to study them in other gauge theories, such as BF and gravitational models. It would also be interesting to analyze the behavior of these systems under the action of the trace operator which, for some Sobolev spaces, provides a consistent and well-defined way to project the dynamics of the bulk onto the boundary \cite{juarez2017boundary}. However, for higher dimensional boundaries, it is important to mention that there are a lot of functional analytic subtleties that have to be taken into account.


%
%
\section*{Acknowledgments}

This work has been supported by the Spanish Ministerio de Ciencia Innovaci\'on y Uni\-ver\-si\-da\-des-Agencia Estatal de Investigaci\'on PID2020-116567GB-C22 grant. Bogar D\'iaz acknowledges support from the CONEX-Plus programme funded by Universidad Carlos III de Madrid and the European Union's Horizon 2020 research and innovation programme under the Marie Sk{\l}odowska-Curie grant agreement No. 801538. Juan Margalef-Bentabol is supported by the AARMS postdoctoral fellowship, by the NSERC Discovery Grant No. 2018-04873, and the NSERC Grant RGPIN-2018-04887. E.J.S. Villase\~nor is supported by the Madrid Government (Comunidad de Madrid-Spain) under the Multiannual Agreement with UC3M in the line of Excellence of University Professors (EPUC3M23), and in the context of the V PRICIT (Regional Programme of Research and Technological Innovation).

%
%
\appendix

\section{Eigen 1-forms problem} \label{apenA}

In this appendix, we study the eigen 1-forms problem \eqref{eigprob2}, i.e.,
\begin{equation} \label{eqfvpa}
\delta \mathrm{d} \vartheta=  \omega^2 \vartheta \quad \textrm{with} \quad  \imath_\partial^*\left(  \ast \mathrm{d}  \vartheta  -  2  \lambda^2 \iota_\nu \ast \vartheta \right)=0 \,.
\end{equation}
Notice that if $\vartheta$ is an eigen 1-form with eigenvalue $\omega^2\neq 0$ and $\upsilon \in \mathcal{E}^1$ [$ \upsilon=\mathrm{d} g$ with $\imath_\partial^*\left(g \right)=0$] then $\langle \upsilon, \vartheta \rangle= \langle \upsilon, \delta \mathrm{d} \vartheta\rangle  / \omega^2=0$, which implies that $\vartheta\in\mathcal{C}^1(\Sigma)\oplus \mathcal{H}^1 (\Sigma)$. In this case, we can use the function $F= \ast \mathrm{d} \vartheta$ to rewrite \eqref{eqfvpa} as
\begin{equation}\label{eqfFla}
 \nabla^2 F = - \omega^2 F \quad \textrm{with} \quad  \imath_\partial^*\left(  F  -  2  \left(\frac{\lambda}{\omega}\right)^2 \iota_\nu \mathrm{d} F \right)=0 \,,
\end{equation}  
which is an eigen \textit{functions} problem. Remember that if we have $F$ then we get $\vartheta$ as $\vartheta= -\ast \mathrm{d} F / \omega^2$ [see \eqref{varphiandF}].

For $\omega = 0$, if we decompose $\vartheta= \vartheta_\textrm{d}+\vartheta_\delta+ \vartheta_\textrm{h}$  and plug it into \eqref{eqfvpa}, we obtain that  $\vartheta_\textrm{d}$ is arbitrary [this sector is not relevant for \eqref{eq:hdeq2comb}], and $\delta \mathrm{d} \vartheta_\delta= 0$ implies $\ast \mathrm{d} \vartheta_\delta=C_1$, where $C_1$ is a constant. Writing $\vartheta_\delta= \delta \phi$ with $ \imath_\partial^*\left( \ast \phi\right)=0$ and defining $f:=\ast \phi$, the equation $\ast \mathrm{d} \vartheta_{\delta}=C_1$ and the boundary condition \eqref{eqfvpa} become
\begin{equation}
\nabla^2  f=-C_1  \quad \textrm{with} \quad \imath_\partial^*\left(  C_1  -  2  \lambda^2 \iota_\nu \left( \mathrm{d} f+\ast \vartheta_\textrm{h}  \right)\right)=0\,. \label{eqfozfa}
\end{equation}
Then, given $\vartheta_\textrm{h}$, we can solve for $f$ in \eqref{eqfozfa} and finally get $\vartheta_\delta$.

\paragraph{The disk}

With the purpose of giving an explicit solution, we restrict ourselves to the case in which $\Sigma$ is a disk of radius $r_0$ (to conform with the conventions of \cite{Bala1994} we make the replacement $\lambda^2 \to  r_0 \lambda^2/2$).

For $\omega \neq 0$ the solutions to \eqref{eqfFla} take the form $ F(r,\theta)= \exp{\left( i M \theta \right)}  J_M(\omega r) $ with $M \in \mathbb{Z}$, and must satisfy the boundary condition
\begin{equation*}
    \imath_\partial^*\left(  F  -  r_0 \left(\frac{\lambda}{\omega}\right)^2 \partial_r F \right)=0  \Rightarrow   J_M(\omega r_0)  -  r_0 \frac{ \lambda^2}{\omega}  J'_M(\omega r_0) =0\,,
\end{equation*}
with $J'_M(x)= \partial_{x} J_M(x)$. These equations give the frequencies: for each $M$ we have a family of $\omega_m$. We denote these infinite (but countable) sets as $\omega_{M, m}$. Then, the eigen 1-form $\vartheta_{M, m}$ with eigenvalue $\omega_{M, m}^2$ is 
\begin{align}
   \vartheta_{M, m}&= \frac{1}{\omega^2_{M, m}} \ast \mathrm{d}  \left(  A_{M, m} \exp{\left( i M \theta \right)}  J_M(\omega_{M, m} r) + A^*_{M, m} \exp{\left( -i M \theta \right)}  J_M(\omega_{M, m} r)\right)\,.
\end{align}
The complex constants $A_{M, m}$ are fixed by the orthonormality condition $ \langle \vartheta_{M n}, \vartheta_{M' n'} \rangle= \delta_{M M'} \delta_{n n'}$. 

For $\omega=0$, we must solve \eqref{eqfozfa}. The solution of $\nabla^2 f =-C_1$ is
\begin{equation*}
    f= -\frac{C_1}{4}r^2+ \sum_{k=1} \left(A_k \cos k \theta +B_k \sin k \theta \right)r^k \,,
\end{equation*}
with $k \in \mathbb{N}$, and $A_k, B_k \in \mathbb{R}$. Using polar coordinates for the harmonic 1-forms $h_k$, and writing $\vartheta_h$ in this basis as $\vartheta_\textrm{h}= \sum_{k=1} \left( c_k h_k + c^*_k \bar{h}_k\right)$ with $c_k \in \mathbb{C}$, the boundary condition in \eqref{eqfozfa} gives
\begin{align*}
    C_1\left(1+ \frac{\lambda^2 r^2_0 }{2}  \right)  -  \lambda^2  \sum_{k=1} &\left( \left(A_k+ \sqrt{\frac{2}{ \pi k}} r_0^{-k} \,\textrm{Im} \,c_{k} \right)k r_0^{k} \cos k \theta\right.\\ &\ +\left.\left(B_k+\sqrt{\frac{2}{ \pi k}} r_0^{-k}\, \textrm{Re} \,c_{k} \right) k r_0^{k} \sin k \theta \right) =0   \,.
\end{align*}
which, as the sine and cosine form an orthonormal basis, implies 
\[A_k=- \sqrt{\frac{2}{ \pi k}}  r_0^{-k} \textrm{Im} \,c_{k}\,, \quad B_k=-\sqrt{\frac{2}{ \pi k}} r_0^{-k} \textrm{Re} \,c_{k}\,,\qquad \mathrm{and} \qquad C_1=0.\]
Therefore, for $\omega=0$, the eigen 1-forms are $\vartheta= \vartheta_\textrm{d}+\vartheta_\delta+ \vartheta_\textrm{h}$, with $\vartheta_\textrm{d}$ and $\vartheta_\textrm{h}=  \sum_{k=1} \left( c_k h_k + c^*_k \bar{h}_k\right)$ arbitrary, and $\vartheta_\delta$ given by
\begin{align*}
   \vartheta_\delta=   \sqrt{\frac{2}{\pi}}\ast \d \left(    \sum_{k=1}  \frac{1}{\sqrt{k}} \Big( \big( \textrm{Im} \,c_{k} \big) \cos k \theta +  \big( \textrm{Re} \,c_{k}\big) \sin k \theta \Big)   \left(\frac{ r}{r_0}\right)^k   \right) \,. 
\end{align*}
Notice that the subspace spanned by the $\vartheta_\textrm{h}$ is infinite dimensional.

\section{The infinite chain of boundary constraints}\label{appendix infinite chain}
In this section, given a form $\alpha$, for $k\in\mathbb{N}$ we denote $\alpha_k:=(\ast\mathrm{d})^k\alpha$, $\alpha_0=\alpha$ and $\alpha_{-k}=0$. We also denote $\pi:=p+\beta\ast A$. Assuming the Hamiltonian dynamics given by \eqref{eq: Hamiltonian}, we have the following easy to prove equation
\[\dot{A}_k=\pi_k\,,\qquad k=1,2,\ldots\]
Using the Gauss constraint $\delta(\pi-2\beta\ast A)=0$ and equations \eqref{eq: Hamiltonian} and \eqref{eq:bcaz1}, it is also straightforward to check that
\[(\ast\pi)_k=-2\beta A_k\,,\qquad k=1,2,\ldots\qquad \mathrm{and} \qquad\dot{\pi}_k=\left\{\begin{array}{lcl}A_{2}+2\beta\ast\pi_0&&k=0\\A_{k+2}-4\beta^2 A_k&\ &k\geq1\end{array}\right. \,.\]
We study now the consistency conditions that arise from equations \eqref{eq:bcaz3} and \eqref{eq:bcanz3}. Notice that they are both of the form $ \Gamma(A)=0$ where $\Gamma=\imath^*_\partial\ast\mathrm{d}$ in the first case and $\Gamma=\imath^*_\partial(\alpha\ast\mathrm{d}+\lambda^2\iota_{\nu}\ast)$ in the second. Notice, however, that the explicit expression of $\Gamma$ is irrelevant for the following argument as long as
\begin{equation}\label{eq: hypothesis dq_per}
     \Gamma(\mathrm{d}A_t)=0\,.
\end{equation}
Applying \eqref{eq:bcaz1}, it is clear that both expressions of $\Gamma$ satisfy \eqref{eq: hypothesis dq_per}.

Under the hypotheses spelled out in the previous paragraph, let us prove that $\Gamma(A)=0$ implies the following infinite chain of boundary constraints:
\begin{align}
    &\Gamma(\pi_0)=0\,,\label{pi=0}\\
    &\Gamma(A_{2k}+2\beta\ast\pi_{2k-2})-4\beta^2\Gamma(A_{2k-2})=0\,,\qquad k=1,2,\ldots\label{boundary1 Gamma}\\
    &\Gamma(\pi_{2k}+2\beta\ast A_{2k})=0\,,\qquad k=1,2,\ldots\label{boundary2 Gamma}
\end{align}

Equation \eqref{pi=0} follows from \eqref{Hemblc3} and
\eqref{eq: hypothesis dq_per}. Equation \eqref{boundary1 Gamma} for  $k=1$ is found by requiring the consistency of \eqref{pi=0} and applying \eqref{Hemblc4}. Equation \eqref{boundary2 Gamma} for $k=1$ is obtained by demanding the consistency of \eqref{boundary1 Gamma} for  $k=1$. Now, assuming that \eqref{boundary1 Gamma} holds for $k$ and \eqref{boundary2 Gamma} holds for $k-1$, we prove that they hold to the next order. First, demanding the consistency of \eqref{boundary1 Gamma} for $k$ leads to
\begin{align*}
    0&=\Gamma\big(\pi_{2k}+2\beta\ast(A_{2k}-4\beta^2A_{2k-2})\big)-4\beta^2\Gamma(\pi_{2k-2})\\
    &=\Gamma\big(\pi_{2k}+2\beta\ast A_{2k}\big)-4\beta^2\Gamma(\pi_{2k-2}+2\beta\ast A_{2k-2})=\Gamma\big(\pi_{2k}+2\beta\ast A_{2k}\big)\,,
\end{align*}
which holds as a consequence of \eqref{boundary2 Gamma} for $k-1$. This proves \eqref{boundary2 Gamma} for $k$. Analogously, demanding the consistency of \eqref{boundary2 Gamma} for $k$ leads to
\begin{align*}
    0&=\Gamma\big(A_{2k+2}-4\beta^2A_{2k}+2\beta\ast \pi_{2k}\big)=\Gamma\big(A_{2k+2}+2\beta\ast \pi_{2k}\big)-4\beta^2\Gamma\big(A_{2k}\big)\,,
\end{align*}
which proves \eqref{boundary1 Gamma} for $k+1$. A final comment is in order now. For $\Gamma=\imath^*_\partial\ast\mathrm{d}$ it is easy to prove that $\Gamma\ast A_k=0=\Gamma\ast \pi_k$ for $k\geq1$. Hence, the infinite chain of conditions simplifies to
\begin{equation}\label{infinite chain lambda 0}
    \left\{\begin{array}{l}
    \Gamma(A_{2k})=0\\
    \Gamma(\pi_{2k})=0
    \end{array}\right.\qquad\equiv\qquad
    \left\{\begin{array}{l}
\imath^*_\partial(\ast\mathrm{d})^{2k+1}A=0\\    \imath^*_\partial(\ast\mathrm{d})^{2k+1}\pi=0
    \end{array}\right. \,.
\end{equation}
%
%

\bibliographystyle{JHEP}
\bibliography{MCS}

\providecommand{\href}[2]{#2}\begingroup\raggedright\begin{thebibliography}{10}

\bibitem{Witten}
E.~Witten, \textit{Quantum field theory and the {J}ones polynomial},
  \href{https://doi.org/10.1007/BF01217730}{\textit{Commun. Math. Phys.}
  {\bfseries 121} (1989) 351}.

\bibitem{elitzur1989remarks}
S.~Elitzur, G.~Moore, A.~Schwimmer and N.~Seiberg, \textit{Remarks on the
  canonical quantization of the {C}hern-{S}imons-{W}itten theory},
  \href{https://doi.org/10.1016/0550-3213(89)90436-7}{\textit{Nucl. Phys. B}
  {\bfseries 326} (1989) 108}.

\bibitem{Wen1992}
X.-G.~Wen, \textit{Theory of the edge states in fractional quantum {H}all
  effects},
  \href{https://doi.org/10.1142/S0217979292000840}{\textit{International
  Journal of Modern Physics B} {\bfseries 06} (1992) 1711}.

\bibitem{wen1995topological}
X.-G.~Wen, \textit{Topological orders and edge excitations in fractional
  quantum {H}all states},
  \href{https://doi.org/10.1080/00018739500101566}{\textit{Advances in Physics}
  {\bfseries 44} (1995) 405}
  [\href{https://arxiv.org/abs/cond-mat/9506066}{arXiv:{\ttfamily
  cond-mat/9506066}}].

\bibitem{Zee:1995avy}
A.~Zee, \textit{{Quantum {H}all fluids}},
  \href{https://doi.org/10.1007/BFb0113369}{\textit{Lect. Notes Phys.}
  {\bfseries 456} (1995) 99}
  [\href{https://arxiv.org/abs/cond-mat/9501022}{arXiv:{\ttfamily
  cond-mat/9501022}}].

\bibitem{Witten:2015aoa}
E.~Witten, \textit{{Three lectures on topological phases of matter}},
  \href{https://doi.org/10.1393/ncr/i2016-10125-3}{\textit{Riv. Nuovo Cim.}
  {\bfseries 39} (2016) 313}
  [\href{https://arxiv.org/abs/1510.07698}{arXiv:{\ttfamily 1510.07698}}].

\bibitem{deser1982three}
S.~Deser, R.~Jackiw and S.~Templeton, \textit{Three-dimensional massive gauge
  theories}, \href{https://doi.org/10.1103/PhysRevLett.48.975}{\textit{Phys.
  Rev. Lett.} {\bfseries 48} (1982) 975}.

\bibitem{DESER1982}
S.~Deser, R.~Jackiw and S.~Templeton, \textit{Topologically massive gauge
  theories},
  \href{https://doi.org/https://doi.org/10.1016/0003-4916(82)90164-6}{\textit{Annals
  of Physics} {\bfseries 140} (1982) 372}.

\bibitem{Balachandran:1994upto}
A.~Balachandran, L.~Chandar and E.~Ercolessi, \textit{Edge states in gauge
  theories: Theory, interpretations and predictions},
  \href{https://doi.org/10.1142/S0217751X95000966}{\textit{International
  Journal of Modern Physics A} {\bfseries 10} (1995) 1969}
  [\href{https://arxiv.org/abs/hep-th/9411164}{arXiv:{\ttfamily
  hep-th/9411164}}].

\bibitem{Asorey2016}
M.~Asorey, A.P.~Balachandran and J.M.~Pérez-Pardo, \textit{Edge states at
  phase boundaries and their stability},
  \href{https://doi.org/10.1142/S0129055X16500203}{\textit{Reviews in
  Mathematical Physics} {\bfseries 28} (2016) 1650020}
  [\href{https://arxiv.org/abs/1505.03461}{arXiv:{\ttfamily 1505.03461}}].

\bibitem{Asorey2013}
M.~Asorey, A.P.~Balachandran and J.M.~P{\'{e}}rez-Pardo, \textit{Edge states:
  topological insulators, superconductors and {QCD} chiral bags},
  \href{https://doi.org/10.1007/jhep12(2013)073}{\textit{JHEP} {\bfseries 2013}
  (2013) } [\href{https://arxiv.org/abs/1308.5635}{arXiv:{\ttfamily
  1308.5635}}].

\bibitem{Agarwal2017}
A.~Agarwal, D.~Karabali and V.P.~Nair, \textit{Gauge-invariant variables and
  entanglement entropy},
  \href{https://doi.org/10.1103/PhysRevD.96.125008}{\textit{Phys. Rev. D}
  {\bfseries 96} (2017) 125008}.

\bibitem{Bala1992}
A.~Balachandran, G.~Bimonte, K.~Gupta and A.~Stern, \textit{Conformal edge
  currents in {C}hern-{S}imons theories},
  \href{https://doi.org/10.1142/S0217751X92002106}{\textit{International
  Journal of Modern Physics A} {\bfseries 07} (1992) 4655}
  [\href{https://arxiv.org/abs/hep-th/9110072}{arXiv:{\ttfamily
  hep-th/9110072}}].

\bibitem{Banados1995}
M.~Ba\~nados, \textit{Global charges in {C}hern-{S}imons theory and the 2+1
  black hole}, \href{https://doi.org/10.1103/PhysRevD.52.5816}{\textit{Phys.
  Rev. D} {\bfseries 52} (1995) 5816}.

\bibitem{Bala2003}
A.P.~Balachandran, S.~Kürk{\c{c}}üoglu and K.S.~Gupta, \textit{Edge currents
  in non-commutative {C}hern-{S}imons theory from a new matrix model},
  \href{https://doi.org/10.1088/1126-6708/2003/09/007}{\textit{JHEP} {\bfseries
  2003} (2003) 007}.

\bibitem{Donnelly:2016auv}
W.~Donnelly and L.~Freidel, \textit{{Local subsystems in gauge theory and
  gravity}}, \href{https://doi.org/10.1007/JHEP09(2016)102}{\textit{JHEP}
  {\bfseries 09} (2016) 102}
  [\href{https://arxiv.org/abs/1601.04744}{arXiv:{\ttfamily 1601.04744}}].

\bibitem{Geiller:2017xad}
M.~Geiller, \textit{{Edge modes and corner ambiguities in 3d
  {C}hern\textendash{}{S}imons theory and gravity}},
  \href{https://doi.org/10.1016/j.nuclphysb.2017.09.010}{\textit{Nucl. Phys. B}
  {\bfseries 924} (2017) 312}
  [\href{https://arxiv.org/abs/1703.04748}{arXiv:{\ttfamily 1703.04748}}].

\bibitem{wen2004quantum}
X.-G.~Wen, \textit{Quantum field theory of many-body systems: from the origin
  of sound to an origin of light and electrons}, Oxford Graduate Texts, Oxford
  University Press, Oxford (2004).

\bibitem{fradkin2013field}
E.~Fradkin, \textit{Field theories of condensed matter physics}, Cambridge
  University Press, Cambridge (2013).

\bibitem{Balachandran:1994up}
A.P.~Balachandran, L.~Chandar and A.~Momen, \textit{{Edge states in gravity and
  black hole physics}},
  \href{https://doi.org/10.1016/0550-3213(95)00622-2}{\textit{Nucl. Phys. B}
  {\bfseries 461} (1996) 581}
  [\href{https://arxiv.org/abs/gr-qc/9412019}{arXiv:{\ttfamily
  gr-qc/9412019}}].

\bibitem{Andrade2005}
T.~Andrade, M.~Ba\~nados, R.D.~Benguria and A.~Gomberoff,
  \textit{$(2+1)$-dimensional charged black hole in topologically massive
  electrodynamics},
  \href{https://doi.org/10.1103/PhysRevLett.95.021102}{\textit{Phys. Rev.
  Lett.} {\bfseries 95} (2005) 021102}
  [\href{https://arxiv.org/abs/hep-th/0503095}{arXiv:{\ttfamily
  hep-th/0503095}}].

\bibitem{Bala1994}
A.~Balachandran, L.~Chandar, E.~Ercolessi, T.~Govindarajan and R.~Shankar,
  \textit{{Maxwell-Chern-Simons electrodynamics on a disk}},
  \href{https://doi.org/10.1142/S0217751X94001357}{\textit{International
  Journal of Modern Physics A} {\bfseries 09} (1994) 3417}
  [\href{https://arxiv.org/abs/cond-mat/9309051}{arXiv:{\ttfamily
  cond-mat/9309051}}].

\bibitem{Park1999}
M.-I.~Park, \textit{Symmetry algebras in {C}hern-{S}imons theories with
  boundary: canonical approach},
  \href{https://doi.org/https://doi.org/10.1016/S0550-3213(99)00031-0}{\textit{Nucl.
  Phys. B} {\bfseries 544} (1999) 377}.

\bibitem{Blasi2010}
A.~Blasi, N.~Maggiore, N.~Magnoli and S.~Storace,
  \textit{{Maxwell{\textendash}Chern{\textendash}Simons theory with a
  boundary}},
  \href{https://doi.org/10.1088/0264-9381/27/16/165018}{\textit{Class. Quant.
  Grav.} {\bfseries 27} (2010) 165018}
  [\href{https://arxiv.org/abs/1002.3227}{arXiv:{\ttfamily 1002.3227}}].

\bibitem{brezis2010}
H.~Brezis, \textit{Functional Analysis, Sobolev Spaces and Partial Differential
  Equations}, Universitext, Springer New York (2010).

\bibitem{Diracnos}
J.F.~Barbero~G., B.~D{\'{\i}}az, J.~Margalef-Bentabol and
  E.J.S.~Villase{\~{n}}or, \textit{Dirac's algorithm in the presence of
  boundaries: a practical guide to a geometric approach},
  \href{https://doi.org/10.1088/1361-6382/ab436b}{\textit{Class. Quant. Grav.}
  {\bfseries 36} (2019) 205014}
  [\href{https://arxiv.org/abs/1904.11790}{arXiv:{\ttfamily 1904.11790}}].

\bibitem{GNH1}
M.J.~Gotay, J.M.~Nester and G.~Hinds, \textit{{Presymplectic manifolds and the
  Dirac–Bergmann theory of constraints}},
  \href{https://doi.org/10.1063/1.523597}{\textit{Journal of Mathematical
  Physics} {\bfseries 19} (1978) 2388}.

\bibitem{Barbero_G_2014}
J.F.~Barbero~G., J.~Prieto and E.J.S.~Villase{\~{n}}or, \textit{Hamiltonian
  treatment of linear field theories in the presence of boundaries: a geometric
  approach},
  \href{https://doi.org/10.1088/0264-9381/31/4/045021}{\textit{Class. Quant.
  Grav.} {\bfseries 31} (2014) 045021}
  [\href{https://arxiv.org/abs/1306.5854}{arXiv:{\ttfamily 1306.5854}}].

\bibitem{margalef2018}
J.~Margalef-Bentabol, \textit{Towards general relativity through parametrized
  theories},  Universidad Carlos III de Madrid.
  \href{https://doi.org/10.48550/arXiv.1807.05534}{\textit{Ph.D. thesis} (2018)
  } [\href{https://arxiv.org/abs/1807.05534}{arXiv:{\ttfamily 1807.05534}}].

\bibitem{HKnos}
J.F.~Barbero~G., B.~D{\'{\i}}az, J.~Margalef-Bentabol and
  E.J.S.~Villase{\~{n}}or, \textit{Generalizations of the {Pontryagin} and
  {Husain-Kuchař} actions to manifolds with boundary},
  \href{https://doi.org/10.1007/JHEP10(2019)121}{\textit{JHEP} {\bfseries 2019}
  (2019) 121} [\href{https://arxiv.org/abs/1906.09820}{arXiv:{\ttfamily
  1906.09820}}].

\bibitem{barbero2021b}
J.F.~Barbero~G., B.~D\'{\i}az, J.~Margalef-Bentabol and E.J.S.~Villase\~nor,
  \textit{{Hamiltonian Gotay-Nester-Hinds analysis of the parametrized
  unimodular extension of the Holst action}},
  \href{https://doi.org/10.1103/PhysRevD.103.064062}{\textit{Phys. Rev. D}
  {\bfseries 103} (2021) 064062}
  [\href{https://arxiv.org/abs/2101.12311}{arXiv:{\ttfamily 2101.12311}}].

\bibitem{Valle-Marc2021}
J.F.~Barbero~G., M.~Basquens, V.~Varo and E.J.S.~Villaseñor, \textit{Three
  roads to the geometric constraint formulation of gravitational theories with
  boundaries}, \href{https://doi.org/10.3390/sym13081430}{\textit{Symmetry}
  {\bfseries 13} (2021) 1430}
  [\href{https://arxiv.org/abs/2109.00472}{arXiv:{\ttfamily 2109.00472}}].

\bibitem{Ferrari1997}
F.~Ferrari and I.~Lazzizzera, \textit{Dirac quantization of the
  {C}hern-{S}imons field theory in the {C}oulomb gauge},
  \href{https://doi.org/https://doi.org/10.1016/S0370-2693(97)00023-3}{\textit{Phys.
  Lett. B} {\bfseries 395} (1997) 250}
  [\href{https://arxiv.org/abs/hep-th/9611211}{arXiv:{\ttfamily
  hep-th/9611211}}].

\bibitem{Segal}
G.~Segal, \textit{Unitary representations of some infinite dimensional groups},
  \href{https://doi.org/10.1007/BF01208274}{\textit{Communications in
  Mathematical Physics} {\bfseries 80} (1981) 301}.

\bibitem{Barbero_G_2016}
J.F.~Barbero~G., J.~Margalef-Bentabol and E.J.S.~Villase{\~{n}}or,
  \textit{Hamiltonian description of the parametrized scalar field in bounded
  spatial regions},
  \href{https://doi.org/10.1088/0264-9381/33/10/105002}{\textit{Class. Quant.
  Grav.} {\bfseries 33} (2016) 105002}
  [\href{https://arxiv.org/abs/1507.05438}{arXiv:{\ttfamily 1507.05438}}].

\bibitem{Jabbari}
M.~Sheikh-Jabbari and A.~Shirzad, \textit{Boundary conditions as {D}irac
  constraints}, \href{https://doi.org/10.1007/s100520100590}{\textit{The
  European Physical Journal C - Particles and Fields} {\bfseries 19} (2001)
  383} [\href{https://arxiv.org/abs/hep-th/9907055}{arXiv:{\ttfamily
  hep-th/9907055}}].

\bibitem{Conner1956}
P.E.~Conner, \textit{The Neumann's problem for differential forms on Riemannian
  manifolds}, vol.~20, American Mathematical Soc., Providence (1956).

\bibitem{abraham1993}
R.~Abraham, J.~Marsden and T.~Ratiu, \textit{Manifolds, Tensor Analysis, and
  Applications}, Applied Mathematical Sciences, Springer, New York (1993).

\bibitem{schwarz2006}
G.~Schwarz, \textit{Hodge Decomposition - A Method for Solving Boundary Value
  Problems}, Lecture Notes in Mathematics, Springer, Berlin, Heidelberg (2006).

\bibitem{watson}
G.~Watson, \textit{A Treatise on the Theory of Bessel Functions}, Cambridge
  Mathematical Library, Cambridge University Press, Cambridge, England (1995).

\bibitem{WEINTRAUB2014361}
S.H.~Weintraub, \textit{{Chapter 8 - de Rham Cohomology}},  in
  \textit{Differential Forms}, S.H.~Weintraub, ed., (Boston), pp.~361--391,
  Academic Press (2014).

\bibitem{juarez2015quantization}
J.F.~Barbero~G., B.A.~Ju{\'a}rez-Aubry, J.~Margalef-Bentabol and
  E.J.~Villase{\~n}or, \textit{Quantization of scalar fields coupled to point
  masses}, \href{https://doi.org/10.1088/0264-9381/32/24/245009}{\textit{Class.
  Quant. Grav.} {\bfseries 32} (2015) 245009}
  [\href{https://arxiv.org/abs/1501.05114}{arXiv:{\ttfamily 1501.05114}}].

\bibitem{juarez2017boundary}
J.F.~Barbero~G., B.A.~Ju{\'a}rez-Aubry, J.~Margalef-Bentabol and
  E.J.~Villase{\~n}or, \textit{Boundary {H}ilbert spaces and trace operators},
  \href{https://doi.org/10.1088/1361-6382/aa65ff}{\textit{Class. Quant. Grav.}
  {\bfseries 34} (2017) 095005}
  [\href{https://arxiv.org/abs/1701.00735}{arXiv:{\ttfamily 1701.00735}}].

\end{thebibliography}\endgroup

\end{document}